\documentclass[a4paper,twocolumn,10pt,accepted=2026-05-03]{quantumarticle}
\pdfoutput=1
\usepackage[numbers,sort,compress]{natbib}
\usepackage{amsmath}
\usepackage{amssymb}
\usepackage{amsfonts}
\usepackage{url}
\usepackage[utf8]{inputenc}
\usepackage[english]{babel}
\usepackage[T1]{fontenc}
\usepackage{graphicx}
\usepackage{bm}
\usepackage{lipsum}
\usepackage{float}
\usepackage{dsfont}
\usepackage{hyperref}
\hypersetup{
    colorlinks=true,
    linkcolor=blue,
    citecolor=cyan,
}

\usepackage{mathtools}
\usepackage[capitalise]{cleveref}
\newcommand{\tr}{\mathop{\rm tr}}
\newcommand{\ket}[1]{\left|#1\right>}
\newcommand{\bra}[1]{\left<#1\right|}

\newcommand{\ttc}{TCI}
\newcommand{\re}[1]{{\color{black}{#1}}}
\newcommand{\textttp}[1]{{#1}} 
\begin{document}
\title{Tensor Cross Interpolation of Purities in Quantum Many-Body Systems}
 \author{Dmytro Kolisnyk}
    \affiliation{Institute of Science and Technology Austria, Am Campus 1, 3400 Klosterneuburg, Austria}
    \orcid{0000-0002-8612-8202}
 \author{Raimel A. Medina}
\affiliation{Institute of Science and Technology Austria, Am Campus 1, 3400 Klosterneuburg, Austria}
    \orcid{0000-0002-5383-2869}
 \author{Romain Vasseur}
\affiliation{Department of Theoretical Physics, University of Geneva, 24 quai Ernest-Ansermet, 1211 Geneva, Switzerland}
    \orcid{0000-0002-4636-4139}
 \author{Maksym Serbyn}
    \affiliation{Institute of Science and Technology Austria, Am Campus 1, 3400 Klosterneuburg, Austria}
    \orcid{0000-0002-2399-5827}

\begin{abstract}

A defining feature of quantum many-body systems is the exponential scaling of the Hilbert space with the number of degrees of freedom. This exponential complexity naïvely renders a complete state characterization, for instance via the complete set of bipartite Renyi entropies for all disjoint regions, a challenging task. Recently, a compact way of storing subregions' purities by encoding them as amplitudes of a fictitious quantum wave function, known as entanglement feature, was proposed. Notably, the entanglement feature can be a simple object even for highly entangled quantum states. However the complexity and practical usage of the entanglement feature for general quantum states has not been explored. In this work, we demonstrate that the entanglement feature can be efficiently learned using only a polynomial amount of samples in the number of degrees of freedom through the so-called tensor cross interpolation (TCI) algorithm, assuming it is expressible as a finite bond dimension MPS. We benchmark this learning process on Haar and random MPS states, confirming analytic expectations. Applying the TCI algorithm to quantum eigenstates of various one dimensional quantum systems, we identify cases where eigenstates have entanglement feature learnable with TCI. We conclude with possible applications of the learned entanglement feature, such as quantifying the distance between different entanglement patterns and finding the optimal one-dimensional ordering of physical indices in a given state, highlighting the potential utility of the proposed purity interpolation method.

\end{abstract}
\maketitle

\section{Introduction}
Recent developments of quantum simulators~\cite{Browaeys2020,Blatt2012,Gross2017,Monroe2021,BlochColdAtoms,Saffman2010} and theoretical methods have dramatically changed our approach to characterizing many-body quantum systems~\cite{PolkovnikovRMP}. In the early days of quantum simulators,  experiments provided access to only a few local observables, often already as ensemble averages. More recently, various experimental platforms gave access to the quantum measurement outcomes of individual quantum degrees of freedom --- full ``snapshots'' of the many-body quantum wave function became available~\cite{Bakr09}. Building on these experimental advances, the shadow tomography framework was proposed~\cite{Elben_2022_review}, that, relying on snapshot taken in different basis,  allows one to efficiently reconstruct general few-body observables, and with additional assumptions on the wave function, even more complicated properties such as global purities~\cite{AFC}.

 Despite these advances, the exponential scaling of the Hilbert space dimension with the number of degrees of freedom remains a fundamental obstacle to the complete characterization of the quantum wave function. While in some cases, quantum states may be simple and easy to characterize, recent experiments realized states emerging from the  dynamics of interacting quantum systems that have non-trivial arbitrary multi-point correlation functions of local observables~\cite{Rispoli}. In contrast to correlation functions, that depend on the choice of operator, entanglement and Renyi entropies allow one to quantify quantum correlations in a basis-independent way for pure states~\cite{Kaufman2016}. For quantum states of interacting systems with local Hamiltonians, entanglement scaling, i.e. the dependence of the bipartite entanglement entropy on the size of region provides valuable insights into the state complexity~\cite{Kaufman2016,SchollwockRMP,AbaninRMP}. In particular, \emph{area-law entangled states} where entanglement stays bounded with the region size in one dimension are considered as ``simple''~\cite{Plenio_AreaLaw}.

Recent numerical progresses in studies of quantum many-body systems were partially fueled by the possibility to \emph{compress} quantum wave functions using an efficient representation. Specifically, for one-dimensional quantum systems matrix product states (MPS) allow for an efficient encoding of area-law entangled quantum states~\cite{SchollwockRMP}, requiring amount of resources that scale polynomially, $O(L\chi^2)$, rather than exponentially in the system size, $L$. Here the parameter $\chi$ is the so-called MPS bond dimension, that determines the rank of the state.  Given the success of the MPS in capturing ground states of gapped and critical systems~\cite{SchollwockRMP}, and describing the early time quantum non-equilibrium dynamics~\cite{Vidal_2004_tebd}, it is natural to ask if MPS can be used for encoding and processing physical properties of the quantum system, such as observables or entanglement.

 Motivated by this question, in the present work we focus on the dataset that consists of \emph{all purities} of a given many-body quantum wave function. This set of information is beyond the standard probes of entanglement or purity scaling, that typically consider only contiguous left/right cuts in the one-dimensional system. Nevertheless, this set of complete purities was considered previously in the literature as entanglement feature~\cite{Akhtar_2020}. Past work has mostly focused on studying the entanglement feature for random Haar states, in part due to its potential uses in shadow tomography scheme~\cite{Akhtar_2023}. Also, the time evolution of the entanglement feature under so-called locally scrambling dynamics was considered~\cite{Kuo_2020}. Notably, in the case of the Haar state, featuring maximal possible violation of area-law entanglement scaling, the entanglement feature that contains information on all bipartite purities turns out to be a remarkably simple object, namely, a matrix product state with the lowest possible nontrivial bond dimension, $\chi=2$~\cite{Akhtar_2020}.

This example of strongly-entangled and complex Haar random state with a simple entanglement feature invites a systematic investigation of the entanglement features of area-law violating states emergent in quantum many-body systems; that is the subject of our work. However, even assuming that such a simple MPS representation of the entanglement feature exists, finding it in practice requires a novel approach beyond those typically used in quantum many-body physics. In particular, the MPS form of the wave function is often found using variational optimization or time propagation -- procedures that do not work for entanglement feature. Finding MPS representations of the entanglement feature by compressing a complete set of purities is highly inefficient, as it requires not only the complete knowledge of the wave function, but also subsequent calculation of all purities.

The key ingredient for computing the entanglement feature proposed in our work is an \emph{application of the tensor cross interpolation algorithm} (\ttc), developed in the context of information processing and compression~\cite{TCI10,Oseledets2011,Savostyanov2011,Savostyanov2014,Dolgov2020,fernandez2024learningtensornetworkstensor}. This algorithm, reviewed in more detail below, constructs the interpolation of the entanglement feature in an iterative way, requiring access only to $O(L\chi^2)$ purity values, where $L$ is the system size, and $\chi$ is an (a priory unknown) rank of the entanglement feature. Hence, the application of \ttc{} to entanglement feature finds its interpolation in a scalable and efficient way (assuming that purities are obtained from experiment~\cite{Kaufman2016} or can be calculated efficiently), provided that it can be represented as a state of the bounded rank $\chi$.

Utilizing the \ttc{} approach to learning and interpolating the entanglement feature, we systematically study entanglement features of various quantum many-body states. In particular, we show that not only the average entanglement feature of Haar random states is representable as an MPS, but individual state realizations drawn from the Haar ensemble also exhibit compressible entanglement features for sufficiently large system size. In addition, we discuss analytical expectations for entanglement features of random MPS states, and numerically study the entanglement feature of critical ground states, highly excited eigenstates~\cite{Sierant_2023}, and other classes of quantum states~\cite{Kitaev,salberger2016fredkinspinchain,Dell_Anna_2016,Movassagh_2016}. Heuristically, we find that  many states violating area-law entanglement scaling have entanglement feature that can be efficiently learned. These results imply that entanglement feature interpolation procedures may be practically useful for characterizing quantum states and processing data from quantum simulators.

Finally, we demonstrate two potential applications of the interpolated entanglement feature. First, we  show that it can be used for quantifying similarity between all purities of different quantum states, thus going beyond usual entanglement scaling considerations. Second, we suggest that entanglement feature may be used to efficiently find the most natural reordering of the spatial indices to minimize the purity of conventional left-right cuts. Such reordering therefore makes entanglement structure more local, potentially allowing one to store the wave function in the form of matrix product states. \re{Problems of similar spirit have been considered before, ranging from tree tensor network structure optimization \cite{Hikihara_2023, Singha_Roy_2020, Singha_Roy_2021} to nontrivial disentangling unitary circuit construction \cite{Hyatt_2017}. We focus specifically on the setting where one is free to change the physical ordering of sites.}

The structure of the remainder of the paper is as follows. First, in Section~\ref{Sec:2} we review the entanglement feature and \ttc{} algorithm that will be used for the interpolation of entanglement feature. Next, in  Section~\ref{Sec:3} we systematically study the feasibility of learning entanglement feature with the \ttc{} approach, starting with analytical results for individual realizations of Haar random states and MPS states, and finishing with numerical study of more general eigenstates of many-body quantum systems. Finally, in Section~\ref{Sec:4} we present applications of entanglement feature to quantify the distance between different quantum states, and also to find the optimal disentangling spatial ordering of degrees of freedom.  \re{The code accompanying numerical part of the study is available at \cite{coderef}.} We conclude with the summary of our results and discussion of open questions in Section~\ref{Sec:5}. Appendices~\ref{App:A}-\ref{APP:ttopt} provide additional details on the considered wave functions and the disentangling algorithm.

\section{Entanglement feature and tensor cross interpolation \label{Sec:2}}

This section introduces in a pedagogical manner two main ingredients of our work: the so-called entanglement feature wave function (Sec.~\ref{Sec:EF-intro})  that encodes all purities of the quantum state, and the tensor cross interpolation algorithm~\cite{fernandez2024learningtensornetworkstensor,TCI10}  (Sec.~\ref{Sec:TCI-intro}) used to approximate the entanglement feature without requiring complete knowledge of all purities. We assume that the reader is familiar with the language of the matrix product states~\cite{Schollw_ck_2011_dmrg}.

\subsection{Encoding purities by wave function \label{Sec:EF-intro}}

\begin{figure}[t]
    \centering
\includegraphics[width=\linewidth]{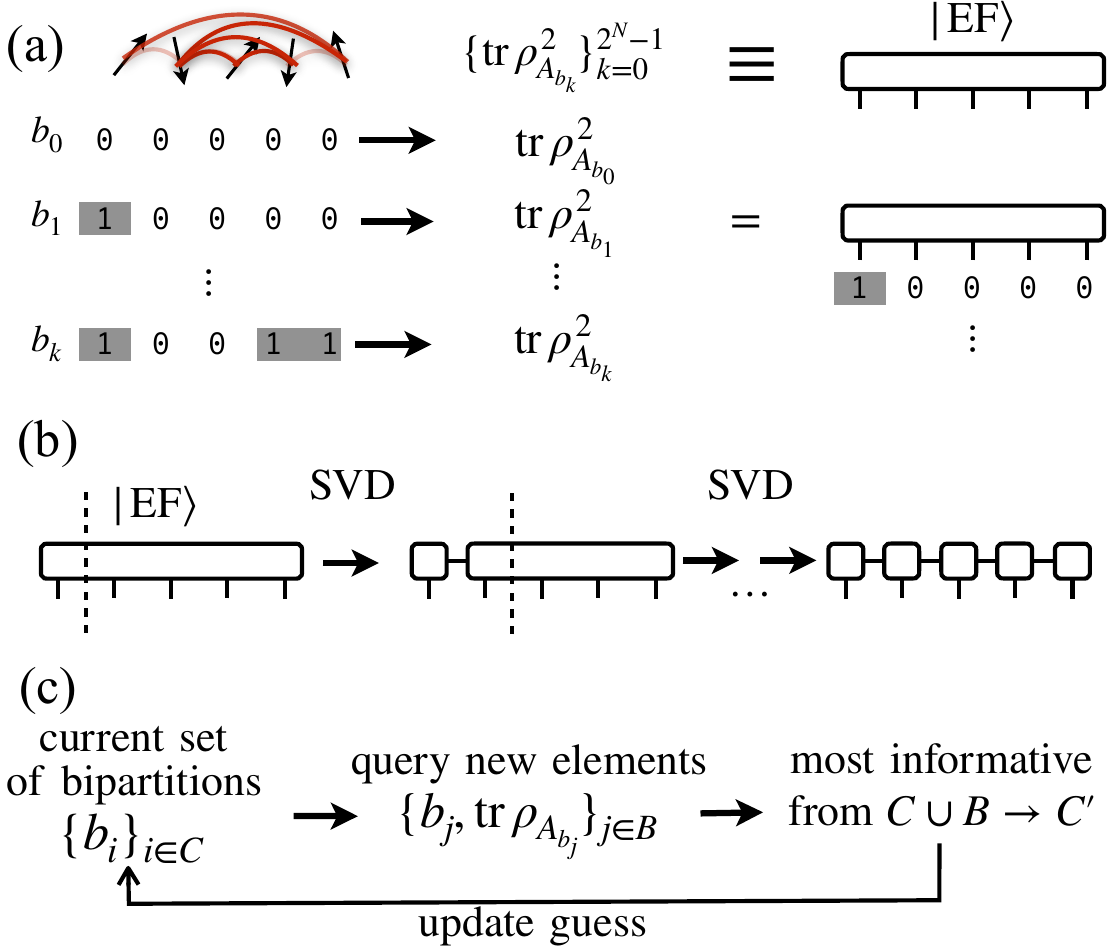}
    \caption{(a) All possible bipartite purities of an entangled quantum state of a many-body system of $L$ spin-1/2 degrees of freedom can be labeled by the bitstrings that specify which degrees of freedom belong to partition $A$ (1-bits) and its complement (0-bits). An exponentially large set of $2^L$ purities for all possible partitions may be encoded in the entanglement feature wave function, $\ket{\rm EF}$, whose amplitudes coincide with the corresponding purities.
    (b) In order to find the matrix product state representation of $\ket{\rm EF}$ one na\"ively needs to first calculate or measure all purities, obtaining the full wave function $\ket{\rm EF}$, and then apply sequential singular value decompositions (SVD).
    (c) The tensor cross interpolation algorithm (TCI)  provides an alternative, where a polynomial in $L$ number of purities is required. The algorithm iteratively determines the best set of purities labeled by the bitstrings $\{b_j\}$ required for interpolating the full $\ket{\rm EF}$ wave function. After convergence, we obtain an \emph{interpolation} of all purities of the system.
    }
     \label{fig:ef_tt_cross_concept}
\end{figure}

A generic pure state $|\psi\rangle$ of a system of $L$ qubits can be characterized by exponentially many bipartite entanglement entropies. The entropies can be stored in an artificial quantum state defining the \emph{entanglement feature}~\cite{Akhtar_2020} of that state:
\begin{equation}
    \label{eq:ef_def}
    \ket{\mathrm{EF}}=\sum_{b\in\{0,1\}^{L}} e^{-S(b)}\ket{b},
\end{equation}
where $S(b)$ denotes an $n$-th Renyi entanglement entropy of the state $|\psi\rangle$ with respect to the bipartite cut into the region $A_b$ [defined by $i\in A_b\Leftrightarrow b_i=1$, see Fig.~\ref{fig:ef_tt_cross_concept}(a)] and its complement. Note that we do not assume the state $\ket{\mathrm{EF}}$ to be normalized. In this paper, we will focus on the second Renyi entropy
\begin{equation} \label{eq:S2-def}
    S(b) = S_2(b) = -\ln \tr \rho_{A_b}^2,
\end{equation}
where $\rho_{A_b}$ denotes the reduced density matrix of the state $|\psi\rangle$ for region $A_b$. For example, the entanglement feature of the state $\ket{\psi}=(\ket{00}+\ket{11})\otimes \ket{0}/\sqrt2$ is given by:
$
\ket{\mathrm{EF}}=1\cdot(\ket{000}+\ket{001}+\ket{110}+\ket{111})+({1}/{2})\cdot(\ket{100}+\ket{010}+\ket{011}+\ket{101})
$, where the bitstrings that have 00 or 11 configuration for the first two bits give purity of 1, and remaining bitstrings where first two spins belong to different regions have purity of $1/2$.

The concept of entanglement feature has proven useful in analyzing the entanglement dynamics in locally scrambled quantum systems~\cite{Kuo_2020, Akhtar_2020} where the quantum dynamics leads to changes in the entanglement that can be followed without reference to the quantum state itself. Moreover, in a locally scrambled dynamics setting, the entanglement feature has found applications in quantum tomography~\cite{Hu_2023, Akhtar_2023} around the classical shadow method~\cite{Huang_2020}. The entanglement feature was also used to study entanglement combined with machine learning methods~\cite{You_2018_ML, Sun_2022}.

From the definition of the entanglement feature given by \cref{eq:ef_def}, we can highlight two important properties: first, for pure states the entanglement feature is $\mathbb{Z}_2$ symmetric, and second, all wave function coefficients are positive numbers. While the latter property is clear from the entanglement feature definition, the first property follows from the invariance of entanglement entropy under the exchange of partition $A$ with its complement, $A^c$. We can resolve this $\mathbb{Z}_2$ symmetry by fixing the first spin to belong to the subregion of interest (fixing $b_1=1$). Then the remaining bit configuration may be specified by the location of the domain walls of the bipartition, thereby corresponding to the standard duality of the Ising model where $b=0,1$ is interpreted as an Ising spin. Applying this duality transformation results in a wave function belonging to a smaller Hilbert space of $L-1$ virtual spins, encoding position of ``domain walls'' separating region $A$ and its complement $A^c$ locally (simple left-right bipartite cut corresponds to a single domain wall, i.e.~bitstring with only one non-zero bit). We call this smaller virtual state space the \emph{dual basis} to distinguish it from the one used in \cref{eq:ef_def}, which we call the \emph{natural basis}. In the dual basis, that will be used in the remainder of the paper if not stated otherwise, the entanglement feature is written similarly to Eq.~(\ref{eq:ef_def}),
\begin{equation}
    \label{eq:ef_def_dual}
    \ket{\overline{\mathrm{EF}}}=\sum_{\bar b\in\{0,1\}^{L-1}} e^{-S(\bar b)}\ket{\bar b},
\end{equation}
and it encodes the same information, however removing the double redundancy stemming from $\mathbb{Z}_2$ symmetry discussed above.

A promising property of the entanglement feature is its ability to \emph{efficiently} encode entanglement entropies of a maximally entangled state \cite{Hayden_2006, Akhtar_2020} (whose entanglement is well approximated by that of Haar random states). In particular, typical maximally entangled states are incompressible, i.e.~they require an amount of memory that is exponential in the system size, $O(2^L)$ to store them. At the same time, it was observed that the complete set of \emph{average purities} of Haar states written as an entanglement feature, results in a much simpler $\ket{\rm EF}$ wave function that can be compressed. Specifically, for one-dimensional systems,  $\ket{\rm EF}$ of typical Haar state is representable as a simple matrix product state (MPS), which can be stored with only $O(L\chi^2)$ resources, where $\chi$ denotes the so-called MPS bond dimension, encoding the maximal rank of the state. While a Haar random state is maximally entangled and complicated, its individual subsystems are just maximally entangled, so the details and local structure of the bipartition do not matter (purity depends only on the size of the subregion), thus explaining the simple structure of the entanglement feature. Inspired by this simplicity, we study a broader class of quantum states relevant for many-body quantum physics using the TCI algorithm, which we review below.

\subsection{Learning MPS states by \ttc{} \label{Sec:TCI-intro}}

Assume that a quantum state has a low-rank entanglement feature state, that can be approximated by a matrix product state (MPS) of low bond dimension, see Refs.~\cite{Schollw_ck_2011_dmrg,cirac2020matrix} for a review of MPS. To obtain an MPS representation of such an entanglement feature in practice, a typical way is to perform a sweep applying singular value decompositions (SVD) over all the sites of the entire entanglement feature tensor, as shown in \cref{fig:ef_tt_cross_concept}(b). Such left-to-right sweep factorizes out a local MPS tensor site after site by (at each step) grouping the leftmost physical index with any virtual ones into a left group of indices, and all the other ones into a right group of indices. This results in an EF tensor reshaped into a matrix to which SVD is applied with a given maximum bond dimension and/or singular value cutoff restriction. However, the application of such SVD sweeping process requires knowledge of \emph{all the elements of entanglement feature}, i.e.~requires the complete knowledge of all purities in advance. This is not scalable, since the number of purity computations (or measurements in experiments) scales exponentially with the system volume.

The scalable TCI approach that allows us to interpolate the full entanglement feature state in the MPS form using only a \emph{limited number of purity computations or measurements} is shown in \cref{fig:ef_tt_cross_concept}(c). It relies on some initial guess of the relevant purities as encoded by the set of bitstrings. Afterwards, the algorithm uses sweep-like iterations to update the set of bitstrings and requires input of the associated purity values.  The required purity values may be calculated numerically on the fly, or provided  by experimental measurements~\cite{Kaufman2016}. The key advantage is that if the entanglement feature is approximated by an MPS of bond dimension $\chi$, the TCI will return a good approximation after requesting calculation only  of an order  $O(L\chi^2)$ purities.

\begin{figure}[t]
    \centering
\includegraphics[width=0.98\linewidth]{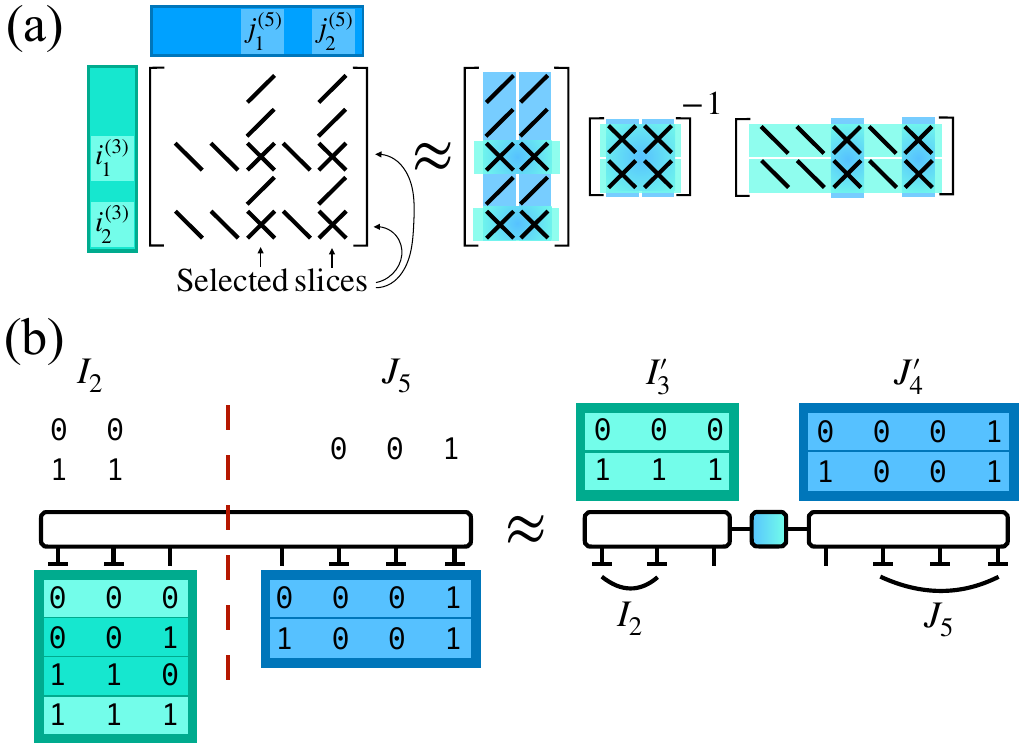}
    \caption{\label{Fig:TCI}
    (a) Schematic depiction of matrix cross interpolation, that selects the set of most relevant rows and columns approximating the entire matrix.  (b) Application of the  matrix cross interpolation during \ttc{} sweep steps. Sets of binary indices to the left (right) of the cut are viewed as multi-indices labeling rows (columns) of the matrix. The indices of the legs adjacent to the cut are not fixed, whereas indices of the underscored legs ($\bot$) are taken from the fixed set, resulting in the overall enlarged index set. Matrix cross interpolation is used to select most relevant indices, resulting in updated index sets.}
\end{figure}

Although the details of the implementation of \ttc{} are not important for the present work, we briefly review its conceptual idea in Fig.~\ref{Fig:TCI}. The \ttc{} algorithm interpolates a tensor from a subset of its entries using the matrix cross interpolation algorithm, which creates a rank-$r$ approximation based on $r$ columns and rows of the matrix (instead of requiring all matrix entries) schematically shown in  Fig.~\ref{Fig:TCI}(a).  The \ttc{} uses matrix cross interpolation as a subroutine, and in a sweeping manner seeks the components of the tensor (labeled by a set of bitstrings) that provide the best interpolation. As a result, the number of purities required for the application of TCI is expected to  grow as $O(L\chi^2)$. \ttc{}, implemented in the \textttp{TensorCrossInterpolation.jl} package~\cite{fernandez2024learningtensornetworkstensor}, can potentially be used in two modes: either to construct an approximation of a chosen bond dimension $\chi$ (mode~I), or to adaptively increase the bond ranks until the error is below the chosen tolerance (mode II). Mode I can be used to create a truncated entanglement feature when there are a limited number of computational resources available. Mode II is perfectly suited to quantify the complexity of the entanglement feature in a given state by calculating the minimum required bond dimension to store the interpolation of entanglement entropies. Therefore, in the remainder of this work, we will adapt the package and use the mode II of TCI algorithm~\cite{fernandez2024learningtensornetworkstensor} to obtain numerical data.

\section{Complexity of the entanglement feature \label{Sec:3}}
This section summarizes our understanding of the entanglement feature complexity for different classes of quantum states. First,  we discuss the behavior of individual realizations of Haar random states and random MPS states. For these two simple classes of states in Sec.~\ref{Sec:EF-Haar}-\ref{Sec:EF-MPS} we obtain analytic insights that are in agreement with numerical results. Finally, in Sec.~\ref{Sec:EF-numerics} we consider the broad class of many-body quantum states and study their entanglement feature numerically using the TCI algorithm.

\subsection{Entanglement feature of Haar states \label{Sec:EF-Haar}}

We use Haar random states as a benchmark to test the applicability of adaptive-mode TCI in evaluating the complexity of quantum state entanglement patterns. For Haar states, the average entanglement entropies can be encoded in a matrix product state with bond dimension $\chi=2$ \cite{Akhtar_2020}, see discussion in Appendix~\ref{App:A} and \cref{eqHaar}. When we consider a single Haar random state (one sample from the distribution), its bipartite entanglement entropies are expected to deviate from the average values by an exponentially small factor. These expectations follow from the mathematical results on the average moments of arbitrary powers of density matrix calculated in Ref.~\cite{Bianchi_2019}. Combining expressions for the average purity and average square of purity, one can show that the variance of the purity of a region $A$ of fixed size, denoted as $\alpha^2$, decays exponentially with the system size $L$
\begin{equation}\label{eq:var}
\alpha^2 \sim  \mathrm{var}\left[\tr\rho_A^2\right]\sim d^{-L},
\end{equation}
where $d$ denotes the local Hilbert space dimension. An alternative argument supporting this scaling can be made using known concentration results for the entanglement entropy of random states~\cite{Hayden_2006}. Such results show that, asymptotically for large system sizes $L$, the probability of observing deviations in entanglement entropy far from the average decreases exponentially with the system's Hilbert space dimension $d^L$ and the deviation considered (squared) $\alpha^2$ with a region-size-dependent prefactor. If we fix a tolerance probability for the entanglement entropy to deviate beyond an in-variance neighborhood of the average value, consistency across different system sizes requires the product $d^L \alpha^2$ to remain constant. This directly implies the expected exponential scaling, $\alpha^2\sim d^{- L}$.

Since the fluctuations of purities decrease with the system size, we expect the entanglement feature of a Haar random state to exhibit a size-dependent transition in terms of complexity. At small system sizes, when sample-to-sample fluctuations in purity are relatively large, the \ttc{} will seek to interpolate the entire set of purities, effectively learning random noise on top of the average value. Since this noise carries no spatial information, we expect that interpolation of the complete set of purities does not allow for any compression and requires a bond dimension that grows exponentially with the system size. In contrast, for sufficiently large systems, when the variance~(\ref{eq:var}) becomes  sufficiently small, \ttc{} working at finite precision, can ignore the fluctations and effectively learn the entanglement feature of the average Haar state, converging to a bond dimension $\chi=2$ MPS.

In order to test these expectations we apply the \ttc{} algorithm to numerically generated wave functions sampled from the Haar distribution. Specifically, we generate Haar random states by sampling unnormalized wave function coefficients from a circularly symmetric complex normal distribution with mean 0 and standard deviation~1, followed by normalization. We process this dense representation to produce purity inputs for the TCI. In all of our tensor cross interpolation tests we use \texttt{crossinterpolate2} function from the julia \textttp{TensorCrossInterpolation.jl}~\cite{fernandez2024learningtensornetworkstensor} package with the following adjusted default parameters: \texttt{tolerance=1e-12}, \texttt{maxiter=4000}, \texttt{maxn\-globalpivot=2}, \texttt{nsearchglobalpivot=2}, where the tolerance parameter bounds the local error allowed during the sweeps.

\begin{figure}[t]
    \centering
    \setlength{\unitlength}{1cm}
    \begin{picture}(7.2,8.9)
        \put(0,4.8){\includegraphics[width=0.4\textwidth]{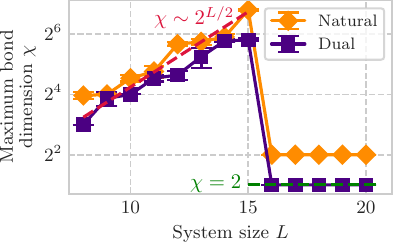}}
        \put(0,0){\includegraphics[width=0.4\textwidth]{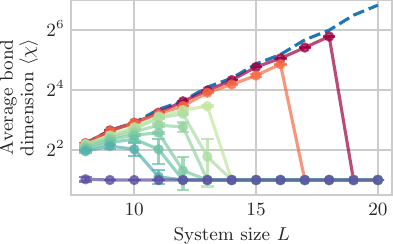}}
        \put(0.2,8.9){(a)}
        \put(0.2,4.2){(b)}
    \end{picture}
    \caption{Tensor cross interpolation of Haar random state entanglement feature.
   (a) Maximal bond dimension as a function of system size for dual and natural basis encoding of entanglement feature. Dual basis encoding gives the expected factor of two  improvement in bond dimension, hence it will be used in the remainder of this work. TCI error threshold~(\ref{eq:error}) was set to $\varepsilon_\text{th} =1.1^{-66}$, which is just below the $0.2\%$.
   (b)~Behavior of  average bond dimension of entanglement feature with the system size shows an error-threshold-dependent transition. The error threshold from left to right curves changes as $\varepsilon_\text{th} = 1.1 ^ {-\{40,50,52,54,56,58,60,70,80\}}$.
   In both panels (a)-(b) the averaging is performed over 10 realizations of Haar random state.
   The error bars denote statistical error across state realizations. The blue dashed curve denotes the maximum possible average bond dimension line.}
    \label{fig:haar_ranks}
\end{figure}

In order to independently assess the performance of \ttc, at every sweep step we define and check the average global relative error,
\begin{equation}\label{eq:error}
\varepsilon = \frac{1}{N_\text{el}}\sum_{b=0}^{N_\text{el}-1}\left|\frac{\mathrm{TCI}[b]-\mathrm{EF}[b]}{\mathrm{EF}[b]}\right|,
\end{equation}
where $\mathrm{EF}[b] = \langle b| \mathrm{EF} \rangle$ is the value of exact entanglement feature, and $\mathrm{TCI}[b]$ is the value of the corresponding interpolation. $N_\text{el}=2^{L-1}$ is the total number of  purities for the pure state defined on $L$ spins.
Although \ttc{} algorithm does not have access to this error, if it falls below a threshold, $\varepsilon\leq \varepsilon_\text{th}$, we stop the algorithm early and note down the TCI bond dimensions. Due to the fact that bond dimension is increased in discrete steps, with increments possibly larger than one, resulting bond dimensions give upper bound on the minimal bond dimension required to achieve the desired tolerance with \ttc{}.

Our numerical TCI tests of randomly sampled Haar states confirm the expectations for the transition in complexity, as shown in \cref{fig:haar_ranks}. First, in \cref{fig:haar_ranks}(a) we compare the maximum bond dimension of the entanglement feature computed in both the natural and dual bases. As expected, the dual basis features lower bond dimension, and both curves display an abrupt transition where $\chi$ sharply drops as the system size increases. \Cref{fig:haar_ranks}(b) studies the sensitivity of this transition to the error threshold described in \cref{eq:error} above. As expected from the analytical results for the variance, increasing the threshold exponentially results in displacement of the transition point to smaller system sizes.

\subsection{Entanglement feature of MPS \label{Sec:EF-MPS}}
\begin{figure}
    \centering
    \includegraphics[width=\linewidth]{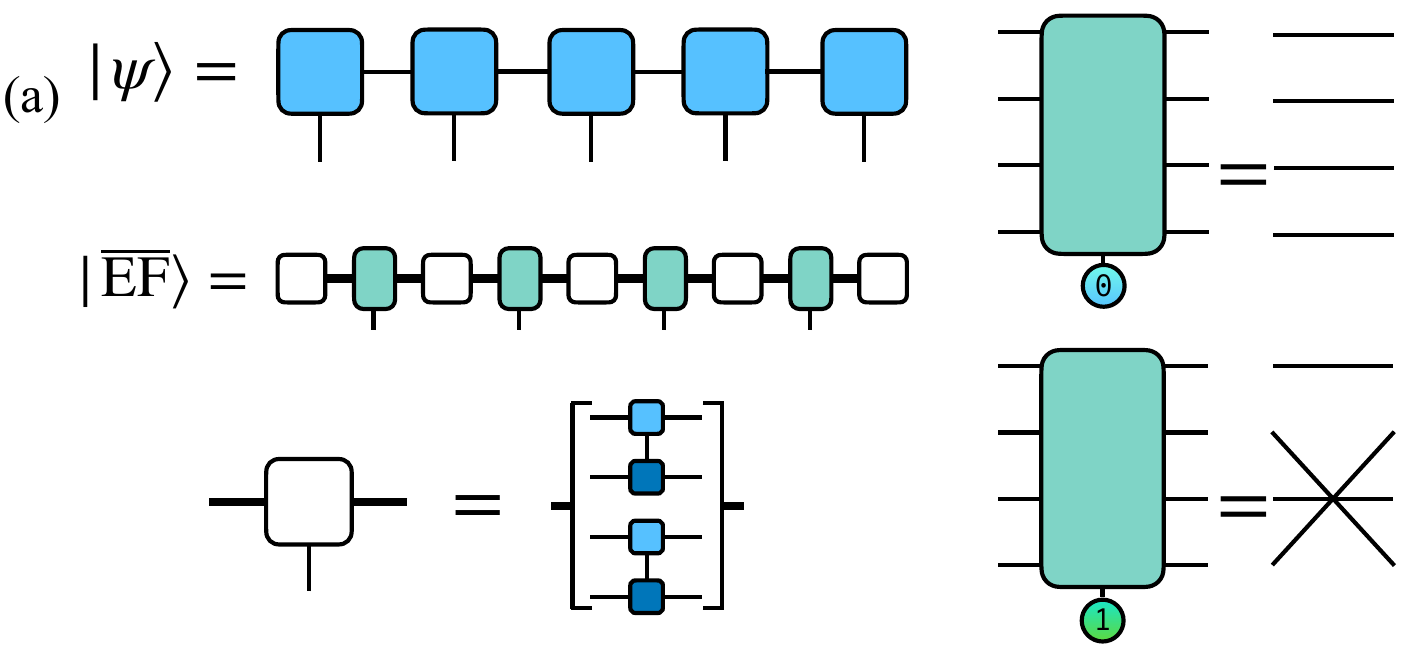}
    \includegraphics{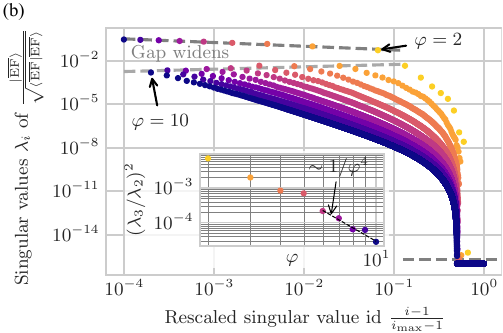}
    \caption{ (a) Construction of the MPS form of the EF in the dual basis from a matrix product state. The tensors of EF MPS are built from 4 copies of original MPS tensors supplemented with the tensor that yields an identity or swap in the space of virtual indices, and the number of physical legs is $L-1$.
    (b) Half-cut entanglement spectrum of the normalized entanglement feature (EF) created from a random MPS of bond dimension $\varphi$ generated on $L=31$ sites. We average squares of singular values ordered by their magnitude, $\lambda^2_1>\lambda^2_2>\ldots$, over 10 random MPS realizations, and plot resulting $\lambda_i$.
    The main feature is the widening of the gap, suggesting that the EF of random MPS with large bond dimension is progressively better described by the $\chi=2$ MPS.}
    \label{fig:es_of_ef}
\end{figure}
Above we discussed the expectations for the EF of the typical Haar random states that are highly entangled, and confirmed these expectations by results of \ttc{} interpolation. In this subsection, we consider another class of tractable states: random matrix product states that are controlled by their bond dimension, denoted as $\varphi$, to distinguish it from the bond dimension of EF MPS encoding. Random MPS states are known to feature nearly maximal possible entanglement, $\ln \varphi$, and in the regime of exponentially large $\varphi$ reproducing Haar random states~\cite{Eisert21}.

The direct construction of the EF MPS representation is obtained from the observation that purity can be calculated using two copies of the density matrix and applying swap or identity operators~\cite{Coser_2014,Kuo_2020}. Moving these operations to the space of virtual indices, we obtain the following compact representation of the EF MPS state shown in Fig.~\ref{fig:es_of_ef}(a). In this figure we defined the tensor with 8 virtual indices and one physical index, that depending on the input of the physical index applies trivial or swap operation to the pair of legs. We note, that although this construction is exact and formally scalable, it results in the rapidly growing bond dimension of the entanglement feature, $\chi=\varphi^4$. In practice, scaling $\chi=\varphi^4$ does not allow to construct entanglement feature in this way when the bond dimension of the original wave function exceeds $\varphi\geq 10$.

At the same time, expectations that at large bond dimension, the random MPS approaches Haar states, suggest that its entanglement feature should have much lower complexity compared to $\chi=\varphi^4$.  Our analytic considerations in  Appendix~\ref{App:A} allow to put these expectations on a more solid basis. By considering the properties of average entanglement feature, we show that when  $\varphi\to\infty$ entanglement spectrum of entanglement feature is expected to have only two relevant singular values, indeed reproducing the case of Haar state. Moreover, we argue that the finite-$\varphi$ corrections are scaling as $\mathcal{O}(\varphi^{-4})$, see \cref{app:ef_of_random_mps} for details.

\begin{figure}
    \centering
    \setlength{\unitlength}{1cm}
    \begin{picture}(7.2,8.9)
        \put(0,4.8){\includegraphics[width=0.4\textwidth]{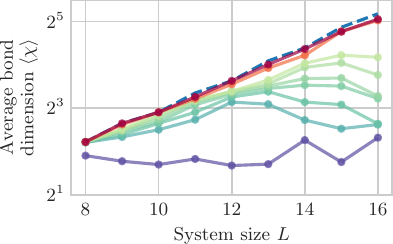}}
        \put(0,0){\includegraphics[width=0.4\textwidth]{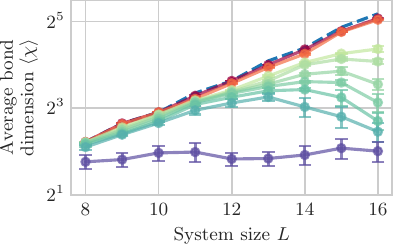}}
        \put(0.2,8.9){(a)}
        \put(0.2,4.2){(b)}
    \end{picture}
    \caption{Average bond dimension of entanglement feature of highly excited eigenstates of (a) TFIM at chaotic point and (b) weakly-disordered TFIM shows a transition similar to the entanglement feature of Haar random states. As the TCI error threshold is changed (bottom to top) in the range $1.1 ^ {-\{40,50,52,54,56,58,60,70,80\}}$, the drop in the average bond dimensions moves to progressively larger system sizes. The blue dashed curve denotes the maximum possible average bond dimension line.}
    \label{fig:ef_complexity_strong_ent}
\end{figure}

The analytic considerations above should be interpreted with caution, since the correct object to consider is the typical entanglement spectrum of entanglement feature, which requires more advanced treatment. Nevertheless, the numerical tests using the direct numerical calculation of the entanglement feature support the analytical results. In Figure~\ref{fig:es_of_ef}(b) we show that as the bond dimension $\varphi$ of the random MPS increases, the gap between the second and the third singular value of the entanglement spectrum of the EF state also grows, leaving only two non-zero singular values in the large bond dimension limit. The inset in Fig.~\ref{fig:es_of_ef}(b) shows that the ratio between third and second singular values squared decays approximately as $1/\varphi^4$ also consistent with heuristic analytical arguments.

Surprisingly, we also see that magnitudes of all other singular values $\lambda_{i\geq 3}$ decay according to an approximate power law with $i$, see \cref{fig:es_of_ef}(b). The origin of this power-law behavior remains to be understood. However, practically it implies that truncating the entanglement feature to the finite bond dimension $\chi<\varphi^4$ incurs an error that decreases polynomially with $\chi$, provided that the singular value decay is sufficiently fast. Finally, we note that random MPS considered here may be viewed as similar to ground states of gapped one-dimensional systems. At the same time, many other interesting examples of MPS states exist, such as MPS approximations to critical ground states, MPS representations of eigenstates of integrable models with long-range correlations, and others. These states substantially differ from random MPS, and hence the present results do not allow to draw conclusion about their entanglement feature.

\subsection{Numerical study of EF complexity \label{Sec:EF-numerics}}

Finally, after describing the two cases where we have some analytical understanding of the EF, we perform numerical study of quantum states relevant in the context of the many-body quantum physics. To this end, we consider two classes of states:  (i) highly entangled, that may be thought to be similar to the Haar random states considered above analytically and (ii) weakly entangled states. In the following, we provide purities, numerically computed from the quantum wave function, into the TCI algorithm operating in the Mode-II (adaptively increasing the bond dimension until certain tolerance) and focus on the dependence of $\chi$ on the system size. To reduce the random influence of pivot initialization in TCI, we average bond dimension results over 10 TCI runs to obtain a single-state TCI data point at a fixed error threshold.

First, as representatives of highly entangled states, we consider highly excited eigenstates (ES) of the chaotic Ising model with transverse and longitudinal field without and with a weak disorder. The Hamiltonian of the system reads,
\begin{equation}
H=J\sum_{i=1}^{L-1}\sigma^{z}_i\sigma^{z}_{i+1}+\sum_{i=1}^{L}\left(g\sigma^{x}_i+(h+r_i)\sigma^{z}_i\right)
\label{eq:def_general_tfim_hamiltonian}
\end{equation}
where in systems with disorder presence we set $J=0.632$, $g=0.902$, $h=0$ and $r_i$ are drawn uniformly from an interval $[0,W]$ where $W$ controls the disorder strength. For the weakly disordered model we use $W=1.0$. We define the Hamiltonian of the chaotic Ising model via \cref{eq:def_general_tfim_hamiltonian} with $J=-1.0$, $g=-1.05$, $h=0.5$, and $W=0$~\cite{Roberts_2015}. We obtain excited states using shift-invert iterations via \texttt{eigs} function (from \textttp{Arpack.jl} julia library) with \texttt{sigma=1e-3} argument to target highly excited energies close to zero energy.

\begin{figure*}
    \centering
    \setlength{\unitlength}{1cm}
    \begin{picture}(18,3.)
        \put(0,0){\includegraphics[width=0.3\textwidth]{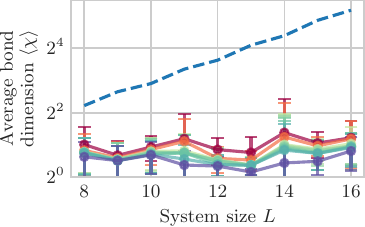}}
        \put(6,0){\includegraphics[width=0.3\textwidth]{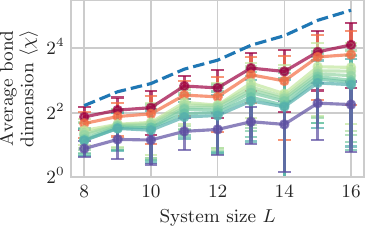}}
        \put(12,0){\includegraphics[width=0.3\textwidth]{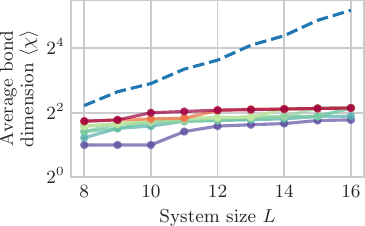}}
        \put(0.2,3.3){(a)}
        \put(6.2,3.3){(b)}
        \put(12.2,3.3){(c)}
    \end{picture}
    \caption{Three qualitatively different behaviors of average bond dimension of entanglement feature of (a) highly excited MBL eigenstates, (b) highly excited MBLT eigenstates, and (c) ground state of critical TFIM. In panel (a) behavior is consistent with bond dimension $\chi$ being constant with system size, in (b) the bond dimension increases exponentially with $L$, and in (c) the data is consistent with the power-law increase. Different colors correspond to TCI error threshold being $1.1 ^ {-\{40,50,52,54,56,58,60,70,80\}}$. The blue dashed curve denotes the maximum possible average bond dimension line.}
    \label{fig:ef_complexity_weak_ent_and_fluctuating_ent}
\end{figure*}
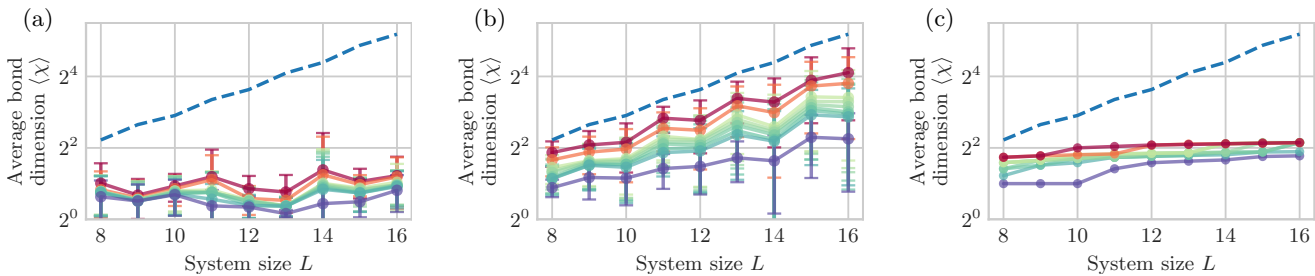

Results of \ttc{} application to extract EF for volume-law entangled states are shown in Fig.~\ref{fig:ef_complexity_strong_ent}(a)-(b). We see that for small system size, the EF bond dimension grows exponentially, transitioning into bond dimension decay approaching a small constant value at large system size that depends on the chosen error, see \cref{fig:ef_complexity_strong_ent}. As the error threshold decreases, the transition point (or area in case of a softer crossover) shifts to a larger system size. This behavior is qualitatively similar to the behavior of random states sampled from the Haar distribution, see Fig.~\ref{fig:haar_ranks} (b). At the same time, the collapse of the average bond dimension in the states reported in \cref{fig:ef_complexity_strong_ent} happens for larger values of error threshold. This suggests that chaotic eigenstates of Ising model feature more entanglement fluctuations compared to the random Haar samples, which is expected due to the local structure of the Ising Hamiltonian.

Second, as representatives of weakly and intermediate entangled states, we consider the three different classes of states. As an example of the states with area-law entanglement scaling we choose the highly excited eigenstates of the disordered Ising model~(\ref{eq:def_general_tfim_hamiltonian}) with parameters $J=0.632$, $g=0.902$, $h=0$, and disorder $W=5$. At this disorder, the model is expected to be in the fully many-body localized phase~\cite{Sierant_2023}, where even highly excited MBL eigenstates are area-law entangled, although their entanglement spectrum decays slower compared to entanglement spectrum of the ground states of gapped Hamiltonians~\cite{Serbyn_2016}. To represent states with intermediate entanglement, we decrease the value of disorder in the Ising model~(\ref{eq:def_general_tfim_hamiltonian}) to $W=2.5$, putting it into finite-size regime of many-body localization phase transition~(MBLT). At this disorder strength, we expect the eigenstates available from finite size exact diagonalization to have more entanglement than area-law~\cite{Alet14,Serbyn_2016,PhysRevB.93.134201,PhysRevLett.119.110604}. Finally, as a third representative of the states with logarithmic entanglement we use the critical Ising model~(\ref{eq:def_general_tfim_hamiltonian}) without disorder and field $W=h=0$ with remaining parameters $J=g=-1$. We obtain the highly excited MBL and MBLT eigenstates in the same manner as described above, while for finding numerically the ground state of critical Ising model we use function \texttt{eigs} from \textttp{Arpack.jl} with parameter \texttt{which=:SR} to target Hamiltonian eigenvalues with the smallest real part.

In the previous Section~\ref{Sec:EF-MPS} we discussed that entanglement feature of generic MPS state with bond dimension $\varphi$ can be also represented as an MPS with bond dimension $\chi =\varphi^4$.  This result implies that states with area-law entanglement, that can be well approximated by an MPS of bond dimension $\varphi$ that is independent of the system size, have entanglement feature with constant bond dimension. Figure~\ref{fig:ef_complexity_weak_ent_and_fluctuating_ent}(a) shows the average bond dimension of numerically constructed entanglement feature of the highly excited MBL eigenstates, which conforms to these expectations. We observe that irrespective of the requested \ttc{} error threshold,  finite bond dimension $\leq 8$  suffices to approximate entanglement feature across the range of accessible system sizes.

Following the same logic, one could conclude that \ttc{} interpolation of entanglement feature of states with logarithmic entanglement scaling should be also efficient, as MPS representation of such states requires bond dimension $\varphi\sim L^a$, leading to expected bond dimension of entanglement feature scaling as $\chi\sim L^{4a}$ in the worst case scenario. This is still parametrically better compared to the exponential scaling of $\chi$ with $L$ expected in the worst case scenario of fully incompressible entanglement feature. Figure~\ref{fig:ef_complexity_weak_ent_and_fluctuating_ent}(b)-(c) shows results of numerical interpolation of entanglement feature of two different classes of states, eigenstates at MBLT and critical ground state of Ising model, revealing more complex picture.

While both eigenstates at MBLT and critical ground state of Ising model have logarithmic entanglement with system size, entanglement feature of MBLT eigenstates in Fig.~\ref{fig:ef_complexity_weak_ent_and_fluctuating_ent}(b) shows exponential increase in average bond dimension $\chi$ with system size. This suggests that the logarithmic average entanglement scaling does not guarantee that entanglement feature,  encoding all possible purities, to be  compressible with bond dimension that scales polynomially in $L$. On the other hand, results for the learning entanglement feature of the critical Ising ground state in Fig.~\ref{fig:ef_complexity_weak_ent_and_fluctuating_ent}(c) fully conform with the expectations discussed above, and show that average bond dimension $\chi$ grows slower than exponentially with the system size $L$ and is not very sensitive to the \ttc{} error threshold.

Beyond the currently presented states, in \cref{APP:models} we additionally study the entanglement feature of more exotic quantum states arising in the context of many-body quantum physics. In particular, we consider the behavior of the entanglement feature of ground state of the Sachdev-Ye-Kitaev (SYK) model \cite{Kitaev}. This state is known to feature volume-law entanglement, and indeed we find that average bond dimension of its entanglement feature grows exponentially with the system size. However, in contrast to Haar random and ETH eigenstates, which display eventual drop in the average bond dimension (see Fig.~\ref{fig:ef_complexity_strong_ent}), we did not observe the similar drop in the bond dimension of entanglement feature of SYK ground state. This suggests that fluctuations of purities of the SYK ground state may have larger magnitude or more complicated structure compared to Haar random states or highly excited eigenstates of chaotic Hamiltonians, so that larger system sizes may be needed to observe decrease in $\chi$.  In addition to SYK, we also studied ground states of the so-called Fredkin~\cite{salberger2016fredkinspinchain,Dell_Anna_2016} as well as the one- and two-color Motzkin chains~\cite{Movassagh_2016}, all known for their area-law violation of entanglement scaling.  While we observe that entanglement feature of Fredkin as well as one-color Motzkin ground state can be learned efficiently, learning purities of two-color Motzkin ground state leads to bond dimension growing exponentially with the system size, see \cref{APP:models} and \cref{fig:ef_complexity_superarea_laws} for details.

To summarize, our numerical experiments with learning entanglement feature of a broad range of quantum states reveal three qualitatively different behaviors: (i)~for Haar-like eigenstates with volume law entanglement, the bond dimension of entanglement feature experiences eventual drop with system size, (ii)~for area-law and some states with sub-volume law entanglement scaling entanglement feature can be captured using $\chi$ constant or polynomially increasing in $L$, and finally the most interesting class of states (iii) where entanglement feature seems to be incompressible. We also emphasize that knowledge of entanglement scaling is not always indicative of the behavior of the entanglement feature, as the latter contains more information.

\section{Applications of Entanglement Feature \label{Sec:4}}
In the previous section we explored the behavior of entanglement feature across a broad range of physical states of quantum many-body systems, finding that in many cases it can be efficiently learned and stored using \ttc{} algorithm. In this section we explore potential applications enabled by the knowledge of entanglement feature. First, in Sec.~\ref{Sec:distance} we use entanglement feature to quantify the distance between entanglement patterns of different quantum states. Second, in Sec.~\ref{Sec:ordering} we demonstrate how the entanglement feature may be used to find the an optimal ordering of spatial indices for decreasing the entanglement scaling of a quantum state. \re{Finally, in Sec.~\ref{Sec:additional-app} we discuss potential broader applications of the entanglement feature.}

\subsection{Distance between entanglement structures \label{Sec:distance}}

Classifying states according to their entanglement content gives valuable insights into existing phases of quantum matter~\cite{PhysRevB.82.155138}, fundamental complexity of quantum operations~\cite{Horodecki2009}, and possibility of state preparation. However, most of the times, a very limited amount of information -- scaling of bipartite entanglement entropy -- is used to infer the entanglement properties of quantum states. The knowledge of entanglement feature, assumed in this section, gives access to a much larger amount of information about purities across all bipartite cuts, and allows us to go beyond just the single-cut scaling analysis.

In order to quantify the distance between entanglement features of different quantum states we use the simple $\ell^2$-norm applied to difference between entanglement features,
\begin{multline}
D\left(\ket{\mathrm{\overline{EF}}_1},\ket{\mathrm{\overline{EF}}_2}\right)=
    \left\lVert\ket{\mathrm{\overline{EF}}_1}-\ket{\mathrm{\overline{EF}}_2}\right\rVert_2
    \\
    = \sqrt{\sum_{\bar b} \left[e^{-S_1(\bar b)}-e^{-S_2(\bar b)}\right]^2},
\end{multline}
that intuitively quantifies the difference between purity of two quantum states over all possible cuts encoded by bitstrings $\bar b$ in the dual basis, see Eq.~(\ref{eq:ef_def}). This overlap can be easily computed using the MPS representation of the entanglement feature. We note that although it is possible to define more complex distances, which can be still efficiently computed using entanglement feature, we focus on the  simplest definition above.

Since some of the states in our catalog are random, we sample the distance matrix $N_\text{samples}=1000$ times by sampling from every random state ensemble generated before (10 random states in Haar, disordered Ising and SYK ensembles, plus 4 states in the GUE-H ensemble) and evaluating the distances for a current realization of states (with TCI error threshold $\varepsilon_\mathrm{th}=10^{-2}$ which did not significantly distort the final EF distance estimates). The compact representation of this distance matrix is available in \cref{APP:raw_dist_data}, \cref{fig:ep_map_distances}. To visualize all pairwise distances we apply stress majorization implemented in \textttp{NetworkLayout.jl} \cite{Stress_2024} to every sampled matrix. We first apply stress majorization to all the non-random states, and fix their positions. Leaving the positions of random states unpinned we apply stress majorization now to all the states across the distance matrix ensemble. In order to better visualize fluctuations among different random realizations, we apply clustering technique, that yields several clusters for each class of random states. Specifically, we cluster the positions of all states using K-means clustering implemented in \textttp{sklearn} python library as \texttt{sklearn.cluster.KMeans} function. We use 3 clusters for the MBLT state (since it was fluctuating the most), 2 clusters for the MBL and GUE-H states, while all other states' positions were significantly less volatile and required no clustering. Every cluster is visualized by its mean and a shaded uncertainty ellipse corresponding to three standard deviations across $N_\text{samples}$ distance matrices.

\begin{figure}[t]
    \centering
        \includegraphics[width=1.0\linewidth]{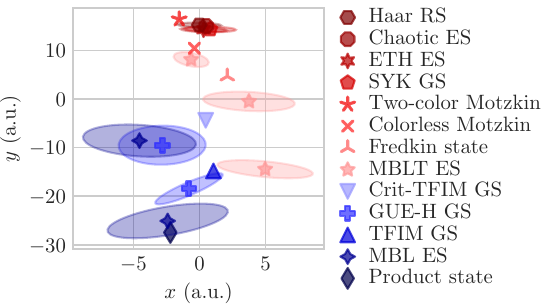
        }
    \caption{Entanglement map illustrating the entanglement-feature-induced distance measure, which is visualized using stress majorization \cite{Stress_2024} of the distance matrix. The legend lists a total of 13 of various states described in the previous section and  \cref{APP:models}. The averaging procedure is described in the main text, and application of clustering to MBLT, MBL and GUE-H states results in multiple representatives for each class of states. System size is fixed to $L=12$.}
    \label{fig:ep_map}
\end{figure}

Despite this somewhat complicated averaging procedure, in the end we obtain an intuitive visualization of distances between different states, referred to as entanglement pattern map in \cref{fig:ep_map}. Intuitively, the visualization implements a two-dimensional projection of the distance matrix with uncertainty (see \cref{fig:ep_map_distances}) that minimizes the stress energy of all state vertices, connected by different springs with the equilibrium distances defined by the distances between entanglement features of states. Hence, although coordinates on the axes in \cref{fig:ep_map} do not have an immediate meaning, proximity of dots implies a relative similarity of EF and vice versa.

\Cref{fig:ep_map} reveals two apparent regions where several classes of states concentrate. The first region is near Haar random state, that also has excited eigenstates of models covered by ETH (chaotic TFIM, weakly disordered TFIM), and slightly further away from Haar state, the SYK ground state. All these states have volume-law entanglement scaling. The second region of concentration encompasses states with area-law entanglement behavior. In particular, next to the product state we see markers corresponding to the mid-spectrum state of the MBL TFIM with disorder $W=5.0$, ground state of the so-called GUE-H model with Hamiltonians generated from a sum of random two-site Hermitian operators, as well as the ground state of the TFIM. Some states with a logarithmic entanglement scaling, such as the ground state of critical TFIM, are also located in the relative vicinity of the product state, but are relatively closer to the Haar cluster, compared to other neighbors of the product state. The other critical states, such as the Fredkin state, and the colorless Motzkin state \cite{salberger2016fredkinspinchain,Dell_Anna_2016,Movassagh_2016,Sugino_2018}, tend to be near the volume-law states, with the more complex states being progressively closer to the Haar state location.

Apart from the two groups of states, we see also apparent outliers in the~\cref{fig:ep_map}, represented by MBLT ES -- a highly excited states at the MBL transition of disordered Ising model with $W=2.5$, as well as the two-color Motzkin. In the case of MBLT state strong fluctuations complicate the entanglement structure of the state, with our clustering method resulting in three separate representatives, all being significantly different from area-law and volume-law clusters in certain realizations. In the case of the two-color Motzkin state, notably, the scaling of the second R\'enyi entanglement entropy is still logarithmic (although the von Neumann entanglement scales as $S_\text{vN}\sim\sqrt{L}$ \cite{Sugino_2018}). At the same time, despite similar scaling of the second R\'enyi entropy, this state is not located between the volume-law and the area-law regions, and goes a bit beyond those regions. The distances between any non-volume-law state and the two-color Motzkin state are larger than those between volume- and non-volume-law states. Hence, we conclude that the distance between entanglement features contains information beyond the entanglement scaling.

\begin{figure*}[t]
    \centering
    \setlength{\unitlength}{1cm}
    \begin{picture}(17,4.)
        \put(0,0.5){\includegraphics[width=0.23\textwidth]{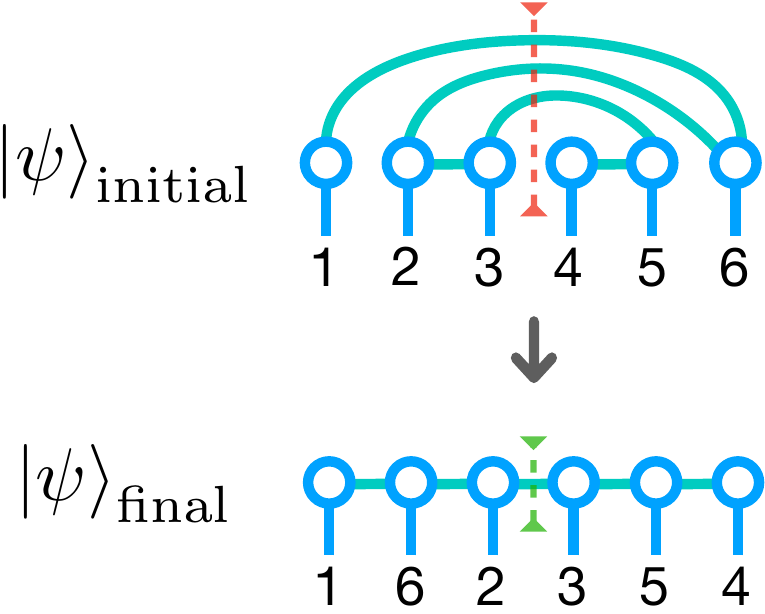}}
        \put(4.4,0){\includegraphics[width=0.34\textwidth]{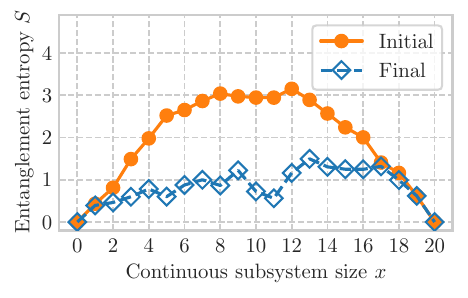}}
        \put(11,0){\includegraphics[width=0.34\textwidth]{{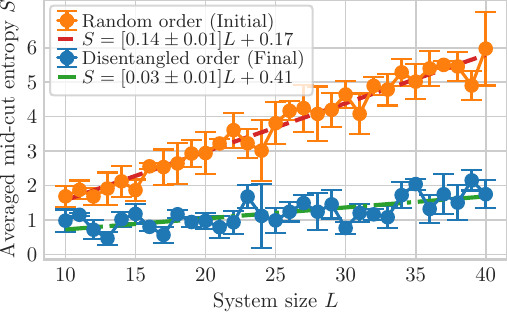}}}
        \put(0.,3.6){(a)}
        \put(4.,3.6){(b)}
        \put(10.5,3.6){(c)}
    \end{picture}
       \caption{
   (a) Seemingly volume-law randomly shuffled MPS can be reordered into a locally clustered area-law state. (b) Disentangler lowers the Page curve of the randomly shuffled MPS state, converting it from a volume-law into an area-law state.
   (c) Half-cut averaged (over 3 random MPS samples) entropy of entanglement for random MPS state with $\chi=3$. Disentangler transforms randomly shuffled MPS state from a volume law state to a close-to-area-law state. Bond dimension 2 was used for the TTOpt procedure. Linear fit coefficient errors come from model error only (spread of entropy values was not taken into account). }
    \label{fig:disentangler_idea}
\end{figure*}
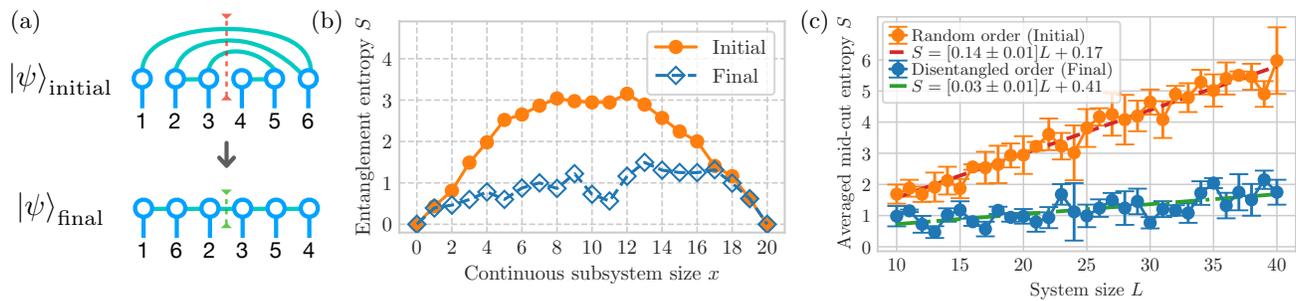

\subsection{Optimal ordering of real space basis \label{Sec:ordering}}

While entanglement and purities are invariant with respect to arbitrary rotations of local Hilbert space basis,  they sensitively depend on the chosen bipartition of the system. Typical characterization of quantum states by their entanglement scaling considers the most natural bipartition, which in turns implies the knowledge of the most natural ordering of the physical sites. While such ordering is obvious in simple cases, for example in one-dimensional spin chains with local interactions, the best ordering of sites in more complex cases, for instance when interactions are represented by a random graph, or in two-dimensional systems, is far from being obvious. In this subsection we demonstrate that the knowledge of entanglement feature allows us to find the close-to-optimal ordering of physical indices that can reduce bipartite entanglement of the state over all left-right cuts.

We test our algorithm on the problem of disentangling an initial area-law entangled MPS, whose physical indices are reshuffled in random (unknown) order. Such reshuffling results in volume-law entanglement scaling of left-right cuts, hence recovering original ordering of indices is expected to result in the drastic reduction of entanglement scaling from volume to area-law, see Fig.~\ref{fig:disentangler_idea}(a).  We assume that we have ``purity oracle'' that returns purities of arbitrary regions upon request, and demonstrate  efficient algorithm that has complexity scaling polynomially rather than exponentially with the system size, $L$. We sample a random MPS of bond dimension $\varphi=3$ using \texttt{randomMPS} function from \textttp{ITensors.jl} julia library~\cite{ITensor} together with a random permutation of its physical indices. We directly build the corresponding EF (as an MPS of bond dimension $\chi=\varphi^4$) and use it as purity oracle.

EF oracle in principle has information about all possible entanglement cuts, however finding the optimal ordering by a na\"ive  search over all permutations scales factorially as $L!$ for a system of size $L$. To avoid this overhead, we use divide-and-conquer approach. We efficiently split the system into two least entangled subsystems of comparable size, assign one of the subsystems to a left set of physical sites (available at the current iteration), the other one to the right, and, finally, repeat the same steps with the obtained smaller subsystems until the smallest possible subsystem of size~$\leq 2$ is reached. The crucial step at the each iteration of the algorithm -- finding the least entangled split over fixed size bipartition -- is performed using an efficient search known as TTOpt~\cite{ttopt}. The idea of the TTOpt algorithm is based on the fact that the maximal modulus element within the maximum determinant-volume submatrix bounds the absolute maximal element within the full matrix \cite{matrixinequalityforttopt}. This means that after efficiently constructing a TCI of a given tensor it is enough to select the maximal element across all the stored cross matrices to obtain a close-to-maximal tensor element (and its indices) at a fraction of a cost a full tensor search.

Application of the TTOpt procedure, however, requires to have a restricted entanglement feature, that encodes only purities corresponding to bipartite cuts of a fixed size. In Appendix~\ref{APP:ttopt} we described a detailed procedure with construction of such restricted entanglement feature, that uses a different basis from the one used in \cref{eq:ef_def,eq:ef_def_dual}. We obtain this restricted entanglement feature using TCI (mode I) that requests purities from the oracle, and interpolates them with a minimal nontrivial bond dimension,  $\gamma=2$, that turns out to be sufficient in practice. After the construction of such restricted entanglement feature with bond dimension $\gamma$, we use the index of the largest element to extract the encoding of the most pure split.

Figure~\ref{fig:disentangler_idea}(b) shows results of the divide-and-conquer disentangling described above with $\gamma=2$ TTOpt, applied to a particular reshuffling of the initial state. We observe that the algorithm substantially decreased entanglement in the center of the reshuffled chain. In order to quantify systematically the gain after the application of the algorithm, we average the results over the ensemble of $N=3$ random initial MPSs. For each fixed system size $L$, we generate $N$ random MPSs and randomly permute their physical indices. After running the disentangler for each state, we obtain two entanglement entropies: one original mid-cut entanglement entropy, and another the final mid-cut one with the site ordering found by the disentangling algorithm. Average entanglement before and after disentangling are compared in \cref{fig:disentangler_idea}(c), which reveals that the algorithm decreases bipartite entanglement by a factor of approximately $3.4$ for largest system size $L=40$. Moreover, initially the entanglement of the bipartite cut in the middle follows volume law, the rate of increase in the bipartite entanglement of the disentangled state is smaller by a factor of approximately $4.4$, and it may be potentially consistent with area-law entanglement already at the $\gamma=2$ TTOpt.

In summary, we demonstrated that information about all purities of the system, that can be efficiently learned for some classes of states using \ttc{} algorithm, may be used to reorder the physical indices of quantum states in order to decrease entanglement of left-right cuts. This may be potentially useful for finding the best index arrangement for MPS representation of the quantum states, in particular for the cases when the obvious ordering of the physical indices is not apparent. \re{More broadly, using entanglement structure to infer geometry was considered earlier \cite{Hikihara_2023, Singha_Roy_2020, Singha_Roy_2021, Hyatt_2017}, and our work provides potential avenue for an efficient algorithm that relies on knowledge of entanglement feature that can be applied in other settings.}

\subsection{Other potential applications\label{Sec:additional-app}}

\re{After illustrating two specific applications of entanglement feature, we outline its potential applications to multipartite entanglement measures and error correction. In contrast to previous sections, here our discussion is more conceptual, and establishing even a proof of concept for these applications is beyond the scope of the present work.}

\re{Before discussing additional applications it is important to discuss an alternative representation of entanglement feature, where coefficients in front of basis states in Eq.~(\ref{eq:ef_def}) are given not by purities (exponential of $n$-th Renyi entropies), but instead directly by the $n$-th Renyi entropy,
\begin{equation}
    \label{eq:lef_def}
    \ket{\mathrm{logEF}}=\sum_{b\in\{0,1\}^{L}} S(b)\ket{b}.
\end{equation}
 From a practical purpose, such \emph{log-entanglement feature} may similarly learned with TCI algorithm, and we observed slightly worse precision compared to learning the conventional entanglement feature. The immediate advantage of such log-entanglement feature is that simple local linear transformation of the basis (MPO with bond dimension one) gives the irreducible log-entanglement feature. Indeed, applying the basis transformation to each local site of log-entanglement feature $\ket{1}=\ket{\tilde 1}$ and $\ket{0}=\ket{\tilde 0}-\ket{\tilde 1}$  maps the two-regions  $\ket{\mathrm{logEF}} = S_i\ket{10}+S_j\ket{01}+S_{ij}\ket{11}$ into $\ket{\widetilde{\mathrm{logEF}}} = S_i\ket{\tilde1 \tilde 0}+S_j\ket{\tilde 0 \tilde 1}+(S_{ij}-S_{i}-S_j)\ket{\tilde 1 \tilde 1}$. We notice that the Renyi entropy of both sites $i$ and $j$ got replaced by their Renyi mutual information.
}

\re{
Above we demonstrated that knowledge of log-entanglement feature allows one to obtain via simple basis transform the multipartite (Renyi) mutual informations. These were recently generalized in Ref.~\cite{santalla2026hyperlinkrepresentationentanglementinclusionexclusion} using notion of entanglement hyperlinks.  Entanglement hyperlinks were introduced as a different way to think about all bipartite entanglement entropies $S_A$, where the idea is to rewrite all $S_A$ in terms of irreducible components $J_A$ (based on set-inclusion-exclusion principle analogy). For example, for a region of three sites $J_{ijk}=S_{ijk}-S_{ij}-S_{ik}-S_{jk}+S_i+S_j+S_k$, so that $S_A=\sum_{I\subseteq A}J_I$, e.g. $S_{ijk}=J_{ijk}+J_{ij}+J_{jk}+J_{ik}+J_i+J_j+J_k$. Relying on the factorization theorem~\cite{santalla2026hyperlinkrepresentationentanglementinclusionexclusion} which states that if a probe region $I$ has nonzero intersections with both the zero-entanglement-entropy region $A$ (with $S_A=0$) and its complement, then $J_I=0$, we can use information obtained efficiently from the log-entanglement feature to estimate how multipartite is the entanglement of the certain state.
}

\re{
In a different direction, we expect that entanglement feature may be also used to provide  insights into error correcting codes~\cite{Terhal,Gottesman} as well as their approximate versions~\cite{Leung,Crepeau,Flammia}. Provided one considers a specific code given by an isometry $V$ mapping from the space of $k$ logical qubits into the space of $p$ physical qubits, one can obtain the purification of the  maximally mixed code state by doubling logical degrees of freedom from $k$ qubits into $k+k$ qubits (with the first copy denoting purifying reference qubits) and considering the action of the map $(\mathds{1}\otimes V)$ on the doubled wave function $\Psi = 2^{-k/2}\sum_{i=1}^{2^k} |b_i\rangle \otimes |b_i\rangle$, where the sum is over the complete set of $2^k$ basis states. Calculating the entanglement feature of quantum wave function $(\mathds{1}\otimes V) \Psi$ will allow for an efficient search for sets of physical qubits whose erasure is  (approximately) correctable, as these are expected to have (approximately) vanishing mutual information with the purifying reference (first $k$) qubits~\cite{Knill,Bravyi}, and also detailed characterization of such sets.  Knowledge of entanglement feature may also potentially assist with spatial rearrangement of qubits along the lines of approach in Sec.~\ref{Sec:ordering} above. Finally, we speculate that entanglement feature may be used to construct the efficiently computable metric for approximate quantum error correcting codes~\cite{Flammia}.
In order to enable such application, however, one should study if entanglement feature of state $(\mathds{1}\otimes V) \Psi$ is efficiently learnable with TCI, which seems to be plausible in view of special entanglement structure of typical error correction codes~\cite{fattal2004entanglementstabilizerformalism,Knill,Bennett,Bravyi2}.
}

\section{Discussion \label{Sec:5}}

In this work, we have demonstrated that the entanglement feature provides an efficient framework for characterizing quantum many-body systems beyond the standard study of entanglement scaling. Our analysis reveals three distinct complexity regimes: strongly-entangled ETH-like states showing a transition from exponential to constant bond dimension with system size, area-law states exhibiting consistently low bond dimension encodings, and states at phase transitions (like MBLT) displaying persistently complex entanglement features.

Learning of the entanglement feature by \ttc{} algorithm and resulting rank of the interpolated state gives conceptual insights into the structure of the complete set of its purities. In addition, we demonstrated two proof-of-concept applications of interpolated entanglement feature. First, we used it to efficiently calculate the distances between complete set of purities for a broad class of quantum states. Among the considered broad class of different quantum states, we observed a natural clustering of states near the Haar random and product state extremes, as well as presence of several outliers that exhibit markedly different entanglement structure. Second, we have also demonstrated that entanglement feature can be used practically for tasks like finding optimal basis orderings that minimize entanglement across bipartitions. This was demonstrated successfully on the test case of randomly shuffled MPS states, where our algorithm recovered area-law scaling from apparent volume-law behavior. \re{We also discussed other, more speculative applications of entanglement feature to multipartite entanglement quantification and quantum error correction in Section~\ref{Sec:additional-app}.}

Potentially interesting further application of the \ttc{} interpolation for purities constructed in our work is the search for resonances within the many-body localized phase~\cite{PhysRevLett.119.110604,PhysRevLett.122.040601}.  These resonances are expected to appear as small clusters of strongly entangled spins among the majority of other spins featuring the area-law entanglement. However, identifying such regions and studying their behavior is a challenging task. The existing algorithm for such resonance search revealed interesting insights~\cite{Herviou_2019}, but its extension to larger system sizes is not possible due  to  exponential scaling of complexity with the system size. Given the success of the \ttc{} learning of the entanglement feature deep in the MBL phase, we expect that it may be used as a basis of the more efficient polynomial-time algorithm for the search and characterization of resonances. When applied to highly excited eigenstates in the MBL phase, or physical states obtained after long-time evolution, such algorithm may reveal useful insights into the resonance structure for large systems.

More broadly, our results raise an interesting question if \emph{other} exponentially large sets of observables in quantum systems can be efficiently interpolated with \ttc{}. The key insight that underlies the successful application of the \ttc{} algorithm to entanglement feature in present work is the fact that purities are invariant under the local basis choice of the quantum wave function.
\re{The TCI algorithm is guaranteed to have no errors in the entries of the entanglement feature that were used as pivots, while all remaining entries are interpolated (see also recent work~\cite{Qin22} bounding the error on the Frobenius norm). Provided, we allow basis rotations, the errors from interpolated elements can spread to all entries in the rotated basis. For the entanglement feature however, basis rotations are not a natural operation since they would mix together purities of different spatial cuts. Therefore the basis-specific nature of the EF is advantageous to keep errors localized.}

Hence, it is interesting to understand if other physically relevant characteristics of many-body eigenstates, such as set of negativities for mixed states~\cite{Horodecki2009}, or set of pairwise mutual informations, featuring similar invariance, may be also efficiently learned via \ttc{} approach. In a different direction, it would be interesting to test the efficiency of the \ttc{} approach for interpolating entire entanglement feature in the restricted setting \re{(which has been efficiently applied to certain quantum many-body perturbative expansions, see the cross-extrapolation reconstruction approach in~\cite{PhysRevB.110.035124})}, for instance allowing it to access purities only of a certain size (or number) of cuts. Such restriction \re{combined with a technique to bound errors~\cite{Qin22} or mitigate noise error post-TCI-learning~\cite{Sakaue2024}} may allow to practically apply the \ttc{} interpolation using data from noisy intermediate scale quantum (NISQ) devices as an input.  \re{Alternatively, approaches that use entanglement feature interpolation from estimated purities~\cite{marko} or state reconstruction~\cite{votto} may provide an alternative avenue for obtaining entanglement feature from NISQ data.}

Achieving progress on the interpolation of exponentially large sets of observables in discrete quantum systems discussed above complements the parallel efforts on application of \ttc{} to describing continuous properties of quantum systems. In particular, the \ttc{} was recently applied to sampling problem of high-dimensional integrals~\cite{Feynman22} and diagrams~\cite{Ishida2024,Murray2024} and representation of functions~\cite{Quantics24,Shinaoka2023,Takahashi2024}, as well as to interpolation of time dependence of local observables~\cite{PhysRevB.110.035124} and correlation functions~\cite{Sakaue2024}~(see also review~\cite{fernandez2024learningtensornetworkstensor} and references therein). We hope that together these efforts will result in more effective description of complex quantum phenomena in equilibrium and out-of-equilibrium interacting quantum systems.

\re{\section*{Code availability}
Code accompanying the study (Ref.~\cite{coderef}) is available at \href{https://github.com/dmytro-kolisnyk/entanglement-feature-learning-tci}{https://github.com/dmytro-kolisnyk/entanglement-feature-learning-tci}.}
\begin{acknowledgments}
We acknowledge useful discussions with Richard Küng on the interpolation methods and error spreading, Ilia A. Luchnikov, Margarita Davydova, and, in particular, Hiroshi Shinaoka, Marc Ritter,  Yuriel Nuñez for useful discussions about TCI and the various workarounds within the \textttp{TensorCrossInterpolation.jl} library.
\re{We also acknowledge the comments of anonymous Referee B, that encouraged us to expand the manuscript with discussion of additional applications of entanglement feature in Section~\ref{Sec:additional-app}.}
M.S. acknowledges discussions with D.~V.~Savostyanov at the 2nd International Quantum Tensor Networks (IQTN) plenary meeting at Flatiron Institute’s Center for Computational Quantum Physics (CCQ) for introduction to the TCI approach.
D.K and M.S. acknowledge support by the European Research Council (ERC) under the European Union’s Horizon 2020 research and innovation program (Grant Agreement No. 850899).
R.V. acknowledges partial support from the US Department of Energy, Office of Science, Basic Energy Sciences, under award No. DE-SC0023999, and the Swiss National Science Foundation (grant 10008234).
This research was supported in part by grant NSF PHY-2309135 to the Kavli Institute for Theoretical Physics (KITP).
\end{acknowledgments}

\appendix
\numberwithin{equation}{section}
\renewcommand{\theequation}{\thesection\arabic{equation}}
\section{Entanglement features of random Haar and MPS states \label{App:A}}

In this appendix, we summarize the analytic expectations for the EF of random Haar and MPS states, using the formalism developed in Refs.~\cite{RTN,RTN2}.

\subsection{EF of random Haar state}\label{app:ef_of_random_haar_state}

Consider the EF of a random Haar state in the thermodynamic limit $L \to \infty$, for a state of qudits. Intuitively, the expectation is that such states are volume law entangled, with a flat entanglement spectrum $S_{n} \simeq (\log d) L_A$ where $L_A$ is the number of spins in the region $A$, where $S_n$ is the $n$-th Renyi entropy. Naively, this would give a product state EF: $\left| \Psi \right. \rangle  = \sum_{\lbrace b_i \rbrace }{\rm e}^{-\log d \sum_i  b_i} \left|  \lbrace b_i \rbrace  \right. \rangle$,
where $b_i = +1$ if site $i$ belongs to region $A$, and $0$ otherwise. This expression is only valid for $L_A \ll L$, and is not ${\mathbb{Z}_2}$ symmetric. To get the full answer, let us compute the EF more carefully, by considering a random state $\rho = \left| \psi  \right. \rangle  \left.\langle  \psi  \right|$ with $ \left| \psi  \right. \rangle  = U \left| 0  \right. \rangle $ and $U \in U(D=2^L)$ a Haar random unitary. Recall that the EF is defined by the purity of this state over all possible partitions
\begin{equation}
\Psi (\lbrace b_i \rbrace) =  {\rm tr} \rho_A^2 = {\rm tr}  \left( {\hat S}_A \rho^{\otimes 2} \right),
\end{equation}
where $ {\hat S}_A$ is the SWAP operation on the two copies of $\rho$ acting in region $A$. For $L\to\infty$, we can estimate the typical behavior of this quantity by averaging it over the random Haar unitary $U$.  Upon averaging, we find that the EF is simply the symmetrized version of the result above
\begin{equation} \label{eqHaar}
 \Psi (\lbrace b_i \rbrace) \sim  {\rm e}^{-\log d \sum_i b_i } +  {\rm e}^{-\log d \sum_i (1-b_i) }.
 \end{equation}
Note that in \cref{eqHaar} the state is just a sum of two product states, or equivalently a sum of two matrix product states with bond dimension one, which can be stored exactly in a bond dimension $\chi=2$ MPS. In other words, even though Haar random states are very complex, the corresponding EF state can be captured by a simple MPS with one bit of entanglement.

\subsection{Random MPS of bond dimension $\varphi$}\label{app:ef_of_random_mps}

Now we consider a random MPS state of bond dimension $\varphi$ with unnormalized density matrix $\rho = \left| \psi  \right. \rangle  \left. \langle  \psi  \right|$. The EF requires computing the purity of this state $ {\rm e}^{-S_2} = {{\rm tr} \rho_A^2 }/{({\rm tr} \rho)^2}$, where the denominator is important to properly define the purity of unnormalized random state. To estimate this quantity, we use the tools developed in Refs.~\cite{RTN,RTN2} to compute entanglement properties of random tensor network states. Properly dealing with the denominator in the definition of the purity requires a replica trick~\cite{RTN2} \footnote{Note that in one dimension it is also possible to go around this normalization problem for random MPS~\cite{PhysRevLett.134.010401,loio2025correlationsspectraentanglementtransitions}, but since we are after nonlinear properties like the typical entanglement spectrum of the EF, a replica trick is needed.}
\begin{equation}
 {\rm e}^{-S_2} = \lim_{n \to 0 }{\rm tr} \rho_A^2 ({\rm tr} \rho)^{n-2} =  \lim_{n \to 0 }{\rm tr} {\hat S_A} \rho^{\otimes n},
\end{equation}
where ${\hat S_A} = \otimes_i {\hat s_i} $ is a permutation operator on the $n$-folded replicated space that acts as SWAP on the first two copies if the site is within the entanglement interval $A$ (and as identity is the site is not in $A$), and as identity on the remaining $n-2$ copies. Since the purity is not a self-averaging quantity, computing the typical behavior of the (entanglement of the) EF goes beyond the scope of this paper. For simplicity, we will gain some intuition by averaging the EF.

Averaging over the random MPS, we find that the (average) EF is given by the partition function of a one-dimensional classical statistical mechanics model, whose degrees of freedom are permutations $g_i \in S_n$, with Hamiltonian~\cite{RTN2}
\begin{equation}\label{Eq:genIsing}
{\cal H}  =  \log \varphi \sum_i (n - C(g_i^{-1} g_{i+1})  + \log d  \sum_i (n- C(s_i^{-1} g_i)).
\end{equation}
Here $C(g)$ is a class function that counts the number of cycles in the permutation $g$. Note this interaction is ferromagnetic, and $C(e) = n$, with $e$ the identity permutation.
Meanwhile $s_i \in S_n $ is a permutation determined by the entanglement partition $\lbrace b_i \rbrace$ corresponding to ${\hat s_i}$ above: if $i \in A$, then $s_i$ is transposition $g_{\rm swap}=(12)$ that swaps copies 1 and 2, that is $s_i$ swaps the first two replicas and leaves the rest intact; whereas if $i \in \overline{A}$, then $s_i=e$ just acts as identity. Intuitively, the term proportional to $\log \varphi$ in Eq.~(\ref{Eq:genIsing}) acts a ferromagnetic interaction, whereas the term proportional to $\log d$ plays the role of a ``magnetic field'' that favors the permutation $s_i$.

For any finite replica number $n$, the resulting EF can be written as an MPS with a finite bond dimension. However, because of the analytic continuation $n \to 0$ in the replica limit, it can have a non-trivial entanglement spectrum. In the limit of the large bond dimension of the random MPS, $\varphi \to \infty$, the ferromagnetic interactions lock all the permutations together, and we recover the result of the random Haar state section above. To see this, note that for $\varphi \to \infty$, we have all permutations become equal to identity $g_i =e$ or the transposition $g_{\rm swap}=(12)$ to minimize the ``magnetic field'' $\log d$ terms. Then using $C(e)=n$ and $C(g_{\rm swap}) = n -1$, the Eq.~(\ref{Eq:genIsing}) reproduces Eq.~\eqref{eqHaar}.

For large but finite bond dimension $\varphi \gg 1$, the leading excitations over uniform permutation $g_i = g_0$ are domain walls with energy cost $\log \varphi$. The minimal excitation is a single domain of a permutation differing from $g_0$ by a single transposition, leading to 2 domain walls, costing energy $2 \log \varphi$. This implies that the finite-$\varphi$ corrections to the Haar case should go as $\sim \varphi^{-2}$: this is indeed consistent with our numerical observations, since generally we expect the EF state corrections of order $\mathcal{O}(\varepsilon)$ to generate entanglement spectrum corrections in the corresponding density matrix only at order $\mathcal{O}(\varepsilon^2)$. Indeed, the half-cut singular value corrections of state can equivalently be thought of as the eigenvalue corrections of the state vector reshaped into a $2^{L/2}\times 2^{L/2}$ matrix. For a generic matrix correction we expect the leading-order eigenvalue corrections to be of the same order as the matrix correction. In this case, the correction matrix is the reshaped EF vector correction, which means that the singular values will be corrected by $\mathcal{O}(\varepsilon)$ terms, leading to $\mathcal{O}(\varepsilon^2)$ entanglement spectrum corrections. For example, consider a state $\ket{\mu_0}=\ket{00}$ with a simple entanglement spectrum $\mathrm{eigs}[\tr_2\rho_{\ket{\mu_0}}]=\{1,0\}$. If we add a generic correction of order $\mathcal{O}(\varepsilon)$ the state becomes $\ket{\mu}\sim \ket{00}+\varepsilon(\alpha_0\ket{00}+\alpha_1\ket{01}+\alpha_2\ket{10}+\alpha_3\ket{11})$, and the previously zero eigenvalue receives a $\mathcal{O}(\varepsilon^2)$ correction, $\mathrm{eigs}[\tr_2\rho_{\ket{\mu}}]=\{1-\alpha_3^2\varepsilon^2,\alpha_3^2\varepsilon^2\}$. This means that a finite-$\varphi$ correction to the Haar-case EF of order $\mathcal{O}(\varphi^{-2})$ should induce a new eigenvalue at order $\mathcal{O}(\varphi^{-4})$, compare with the inset of \cref{fig:es_of_ef}. We leave a detailed theoretical characterization of the entanglement spectrum of the EF of typical random MPS for future work.

\begin{figure*}[t]
    \centering
    \setlength{\unitlength}{1cm}
    \scalebox{0.95}{\begin{picture}(18,7.2)
        \put(0,4){\includegraphics[trim={0 10 0 0},clip]{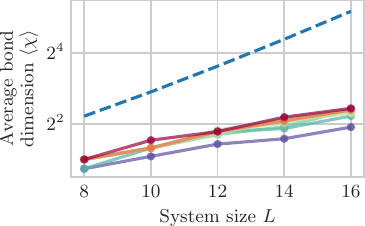}}
        \put(6.5,4){\includegraphics[trim={21 10 0 0},clip]{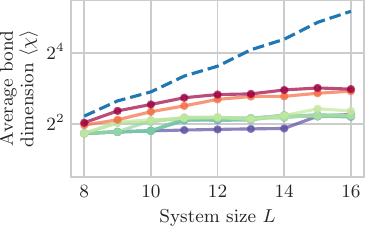}}
        \put(12.4,4){\includegraphics[trim={21 10 0 0},clip]{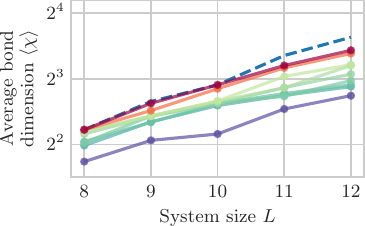}}
        \put(0,0){\includegraphics[trim={0 0 0 0},clip]{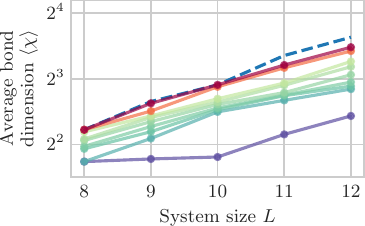}}
        \put(6.5,0){\includegraphics[trim={21 0 0 0},clip]{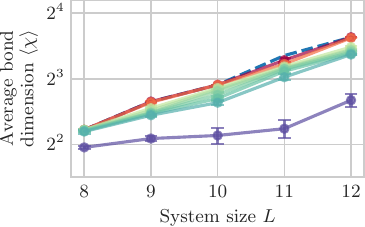}}
        \put(12.4,0){\includegraphics[trim={21 0 0 0},clip]{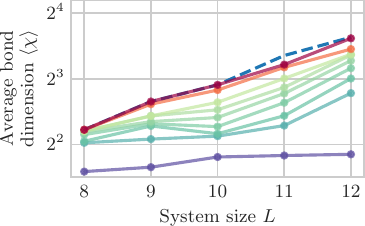}}
        \put(0.7,7.5){(a)}
        \put(6.4,7.5){(b)}
        \put(12.3,7.5){(c)}
        \put(0.7,3.85){(d)}
        \put(6.4,3.85){(e)}
        \put(12.3,3.85){(f)}
    \end{picture}}
    \caption{EF complexity of (a) Fredkin state, (b) colorless Motzkin state, (c)-(d)  two-color Motzkin state considering standard entanglement feature  in (c) and also entanglement feature defined using von Neumann entropy instead of $S_2$ in panel (d), (e)~EF of individual realizations of the SYK ground states (10 state samples), (f) EF created from average entanglement entropies of the SYK ground states (10 state samples). Different colors correspond to TCI error threshold being $1.1 ^ {-\{40,50,52,54,56,58,60,70,80\}}$. The blue dashed curve denotes the maximum possible average bond dimension line.}
    \label{fig:ef_complexity_superarea_laws}
\end{figure*}
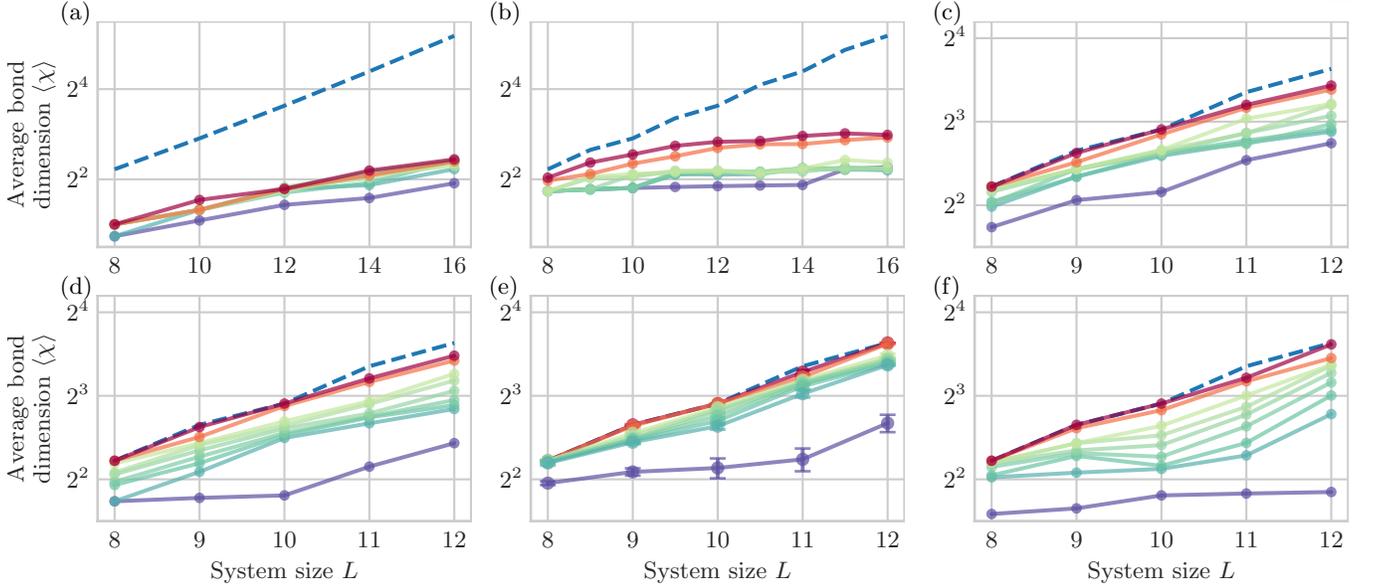
\begin{figure*}[t]
    \centering
    \setlength{\unitlength}{1cm}
    \scalebox{0.98}{\begin{picture}(18,3.4)
        \put(0,0){\includegraphics[width=0.3\textwidth]{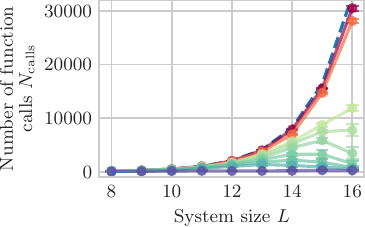}}
        \put(6,0){\includegraphics[width=0.3\textwidth]{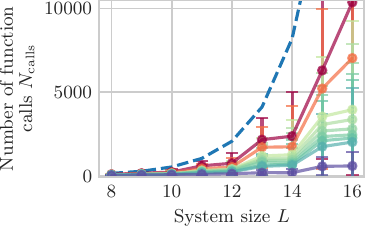}}
        \put(12,0){\includegraphics[width=0.3\textwidth]{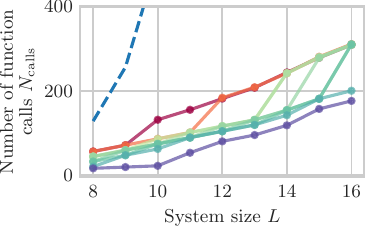}}
        \put(0.2,3.3){(a)}
        \put(6.2,3.3){(b)}
        \put(12.2,3.3){(c)}
    \end{picture}}
    \caption{Number of queries to the purity function used by TCI for the Mode-II reconstruction of (a) highly excited eigenstates of the weakly-disordered TFIM, (b) highly excited MBLT eigenstates, and (c) ground state of critical TFIM. In panel (a) behavior mirrors the bond dimension behavior growing exponentially along the system sizes that were difficult to interpolate (compare with \cref{fig:ef_complexity_strong_ent}), in (b) the number of queries is lower but grows steadily at a rate consistent with exponential or polynomial growth depending on the error threshold (compare with \cref{fig:ef_complexity_weak_ent_and_fluctuating_ent}) and in (c) the number of queries is astonishingly small and grows at a slow linear pace (compare with \cref{fig:ef_complexity_weak_ent_and_fluctuating_ent}). Different colors correspond to TCI error threshold being $1.1 ^ {-\{40,50,52,54,56,58,60,70,80\}}$. The blue dashed curve indicates the maximum possible number of purity function calls.}
    \label{fig:ef_num_calls}
\end{figure*}
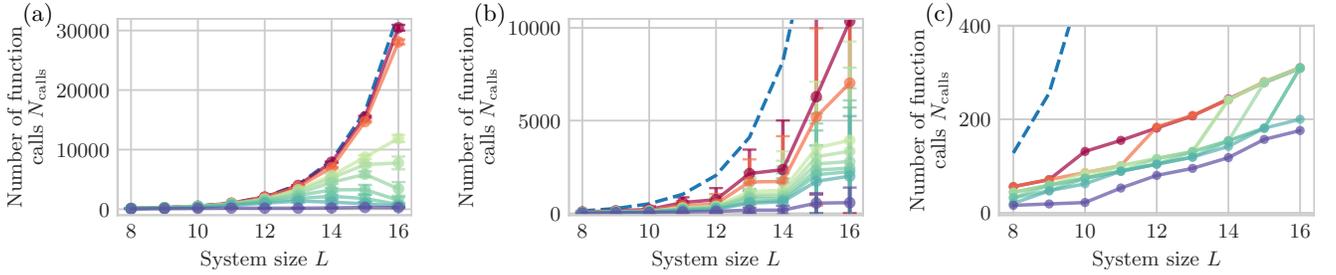

\section{Additional many-body quantum states}\label{APP:models}
Here we discuss the additional quantum states that were only briefly mentioned in the main text, present the scaling of EF bond dimension for these states, discuss the query efficiency of the TCI reconstruction, and also show the raw distances between all the entanglement features.

\begin{figure*}[t]
    \centering
\includegraphics[width=.99\linewidth]{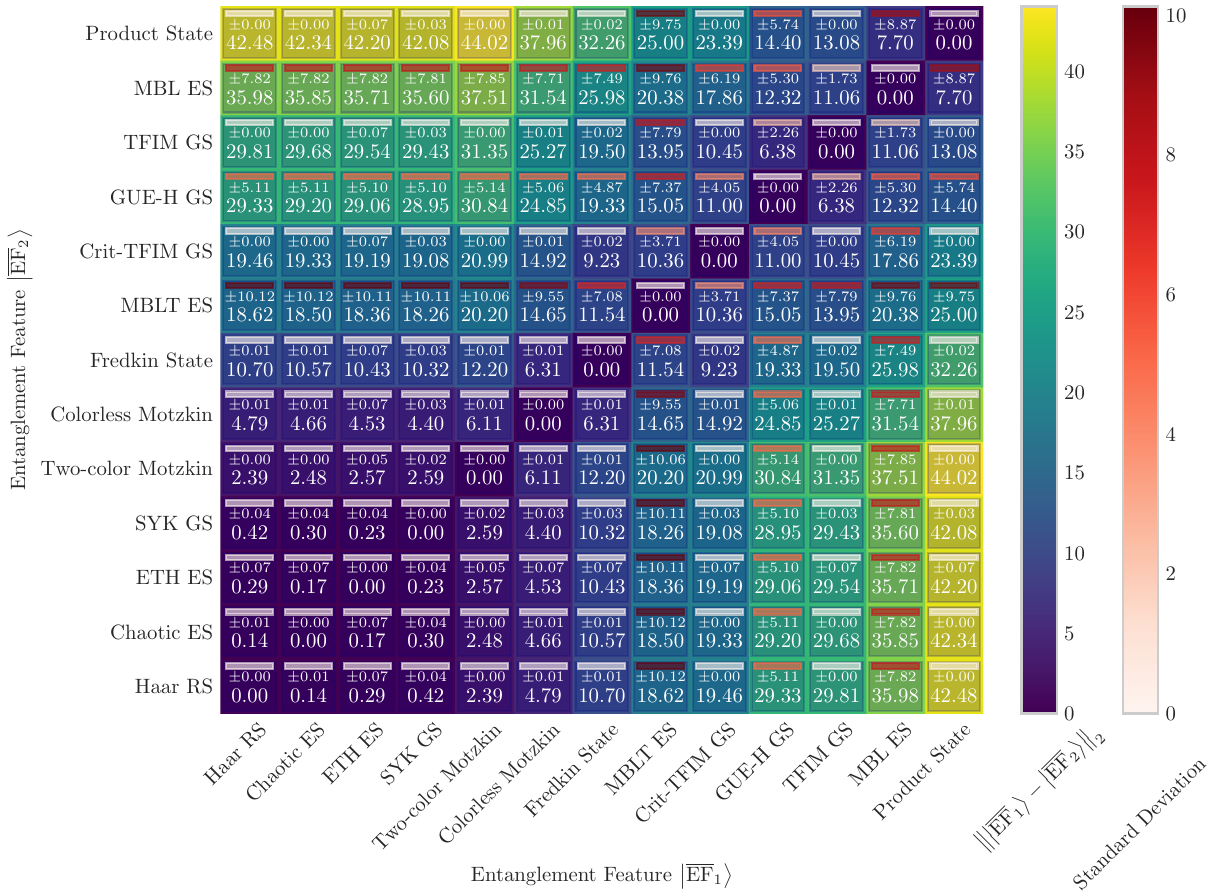}
    \caption{Raw statistics of the entanglement feature distances. System size is fixed to $L=12$.}
    \label{fig:ep_map_distances}
\end{figure*}

\subsection{SYK model and partial tracing in fermionic systems}
\label{APP:fermionic_trace}
The SYK GS state is the ground state of the SYK model \cite{Kitaev}. The ensemble of Hamiltonians we consider is:
\begin{equation}
    H_\mathrm{SYK}=-\frac{\sqrt{6}}{N^{3/2}}\sum_{1\leq i<j<k<l\leq L}4J_{ijkl}\chi_i\chi_j\chi_k\chi_l
    \label{eq:Hsyk}
\end{equation}
where $J_{ijkl}$ are independent mean zero, variance one Gaussian random variables, and $\chi_{i}$ are Majorana fermions. We study the model numerically by grouping Majorana operator pairs $\chi_{2k-1},\chi_{2k}$ into the usual Dirac fermion operators $c_k,c^\dag_k$ and use those Dirac fermion sites to define a bipartition of system (without splitting Majorana pairs into different regions). For the fermionic computations we use the canonically ordered orthonormal basis:
\begin{equation}
\{\ket{i_1 i_2\ldots i_L}\}=\{(c_1^\dag)^{i_1}(c_2^\dag)^{i_2}\ldots (c_L^\dag)^{i_L}\ket{\Omega}\},
\label{eq:c_basis}
\end{equation}
with each index taking only two possible values, $ i_{k{=1,\ldots,L}}\in\{0,1\}$.
To create the Hamiltonian matrix numerically in the dirac fermion basis from \cref{eq:c_basis} we encode majorana operators as follows (for $k=1,\ldots,L$):
\begin{align}
    \sqrt{2}\chi_{2k-1}&=c^\dag_k+c_k\\
    \sqrt{2}\chi_{2k}&=i(c_k-c^\dag_k)
\end{align}
To compute entanglement entropy numerically we require the partial tracing procedure for the full density matrix in the basis specified by \cref{eq:c_basis}. We follow the definitions from \cite{Vidal_2021} where the authors start by defining the partial trace of the last mode:
\begin{multline}
    \tr_L((c_1^\dag)^{i_1}\ldots (c_L^\dag)^{i_L}\ket{\Omega}\bra{\Omega}(c_L)^{j_L}\ldots (c_1)^{j_1})=\\=\delta_{i_Lj_L}(c_1^\dag)^{i_1}\ldots (c_{L-1}^\dag)^{i_{L-1}}\ket{\Omega}\bra{\Omega}(c_{L-1})^{j_{L-1}}\ldots (c_1)^{j_1}
    \label{eq:lastsitetraceout}
\end{multline}
Combining \cref{eq:lastsitetraceout} with canonical anticommutation relations $\{c_j,c^\dag_k\}=\delta_{j,k}$ results in a definition for an arbitrary mode:
\begin{multline}
    \tr_m((c_1^\dag)^{i_1}\ldots (c_L^\dag)^{i_L}\ket{\Omega}\bra{\Omega}(c_L)^{j_L}\ldots (c_1)^{j_1})=\\=(-1)^{{\sigma}}\delta_{i_mj_m}(c_1^\dag)^{i_1}\ldots(c_{m-1}^\dag)^{i_{m-1}}(c_{m+1}^\dag)^{i_{m+1}}\ldots (c_{L}^\dag)^{i_{L}}\ket{\Omega}\\\bra{\Omega}(c_{L})^{j_{L}}\ldots (c_{m+1})^{j_{m+1}}(c_{m-1})^{j_{m-1}}\ldots(c_1)^{j_1}
\end{multline}
where the extra sign $\sigma$ comes from the anticommutation of mode $m$ to the last site:
\begin{equation*}
    \begin{dcases*}
        \sigma=0 & if  $i_m=j_m=0$,  \\
        \sigma=\sum_{q=m+1}^L{}(i_q+j_q) & if $i_m=j_m=1$.
    \end{dcases*}
\end{equation*}
Using such tracing procedure iteratively we calculate the reduced density matrix of the required bipartition and purity, required for the interpolation of the entanglement feature.
\subsection{Motzkin wave functions}

Now we briefly review additional area-law violating states. We start from a $c$-color Motzkin state. This state is defined on a space with local spin-$c$ degrees of freedom. It is an equal-weight superposition of color-balanced Motzkin paths \cite{Sugino_2018,Movassagh_2016}. Such path is a sequence of steps directed up $(1,1)$, down $(1,-1)$ or sideways $(1,0)$ that ends on the same height ($y$-coordinate) as the starting position, and never goes below the starting height. The final crucial condition of being a color-balanced path is that the colors of the up and down steps on the same height level should match (this condition creates strong correlations between distant spins, leading to large entanglement). Throughout the paper we considered a two-color and a one-color (or equivalently, colorless) Motzkin state. As for the Fredkin state label we used earlier, it was actually referring to the spin-$\frac{1}{2}$ Fredkin state, where in general Fredkin states are half-integer spin-$(c-\frac{1}{2})$ analogues of $c$-color Motzkin states. They are also equal-weight superpositions, but of color-balanced Dyck paths, which are Motzkin paths without the sideway $(1,0)$ steps~\cite{Sugino_2018,salberger2016fredkinspinchain,Dell_Anna_2016}.
\subsection{GUE-H states}
We also consider ground states of a random spin-$\frac{1}{2}$ Hamiltonian constructed as a sum
\begin{equation}
    H_\mathrm{GUE}=h_{1;2}+h_{2;3}+\ldots+h_{L-1;L}
\end{equation}
of two-site random local terms $h_{j;k}$ drawn from the Gaussian Unitary Ensemble (GUE). These states model ground states of local Hamiltonians, and are labeled as GUE-H GS on the entanglement pattern map.
\subsection{Additional TCI complexity curves and raw entanglement feature distance data}\label{APP:raw_dist_data}
For completeness we add the entanglement feature complexity curves for the rest of the area-law violating states that are briefly discussed in the main text, see \cref{fig:ef_complexity_superarea_laws}. We also clarify the intuition that the bond dimension of the tensor-cross-interpolated EF closely follows the number of EF elements accessed during the TCI procedure in \cref{fig:ef_num_calls}. Moreover, we also show the raw averages of distances between entanglement features of all pairs of states in~\cref{fig:ep_map_distances}. Single realizations of this data were used to obtain the Fig.~\ref{fig:ep_map} in the main text.

\section{Restricted discrete optimization}\label{APP:ttopt}
In one of the EF applications, when looking for a minimal-entanglement 1D arrangement of a quantum state in Sec.~\ref{Sec:ordering}, an important subroutine that we utilize finds a close-to-weakest entangled bipartition among fixed size ones. To implement this, we choose a fixed-subsystem-size basis that, given a subsystem size $|A|$, specifies a bipartition of such size by specifying the position of the first spin belonging to $A$, then specifying the position of the second spin from $A$ chosen from the leftover $L-1$ possible positions, and so on until the position of the $|A|$-th spin is specified. Note that different basis elements may refer to the same bipartition, and that the local dimensions are $L,L-1,\ldots,L-|A|+1$. For example, in a system with $L=7$ spins a basis element $(4,6,2,2)$ refers to the bipartition with a subsystem $A=\{4,7,2,3\}$. Finding a maximal element of a tensor that stores purities in the fixed-subsystem-size basis is equivalent to performing restricted-subsystem-size search for the maximal EF element, which is our goal. Therefore, it is enough to consider the problem of maximizing a discrete purity function given implicitly by on-demand generated entries from EF in the fixed-subsystem-size basis. We solve this problem by using the TTOpt algorithm~\cite{ttopt}, that creates a TCI of the  oracle function (which in our case is the purity oracle) and then simply selects the largest-modulus element among all pivot matrix entries. This approach is motivated by the fact that the maximum-modulus element in a maximum-determinant-volume submatrix bounds the maximum-modulus element in the full matrix \cite{matrixinequalityforttopt}, and by empirical tests \cite{ttopt}.

%

\end{document}